\documentclass[review,3p,times]{elsarticle}
\usepackage{amsmath,hyperref}
\usepackage{lineno}
\usepackage{color}
\usepackage{subfigure}
\usepackage{epsfig}
\usepackage{caption,xcolor,float,graphicx}
\usepackage{graphics}
\usepackage{epstopdf}
\biboptions{sort&compress}
\journal{Applied Thermal Engineering}

\begin{document}

\begin{frontmatter}

\title{Thermocapillary migration of a self-rewetting droplet on an inclined surface: A phase-field simulation}	
\author[mymainaddress]{He Yan}

\author[mymainaddress]{Lei Wang\corref{mycorrespondingauthor}}
\cortext[mycorrespondingauthor]{Corresponding author}
\ead{wangleir1989@126.com}
\address[mymainaddress]{School of Mathematics and Physics, China University of Geosciences, Wuhan 430074, China}

\author[mysecondaddress]{Jiangxu Huang}
\address[mysecondaddress]{School of Mathematics and Statistics, Huazhong University of Science and Technology, Wuhan 430074, China}

\author[mythirdaddress]{Yuan Yu}
\address[mythirdaddress]{School of Mathematics and Computational Science, Xiangtan University, Xiangtan 411105, China}
	
\begin{abstract}
In this paper, we investigated the thermocapillary migration of a self-rewetting droplet on an inclined surface using a phase field based lattice Boltzmann method. Unlike the normal fluid whose surface tension decreases linearly with temperature, the self-rewetting fluid consider in the current work has a quadratic temperature dependence of surface tension with a well-defined minimum. we first explored the influence of the Marangoni number on droplet migration, and found that the droplet hardly deforms and migrates slowly when the Marangoni number is small. However, as the Marangoni number increases, the droplet begins to deform and elongate, and its migration speed increases. Subsequently, we studied the effect of surface wettability on droplet migration. The results show that the droplet migrate towards regions of higher surface energy on hydrophilic surfaces and in the opposite direction on hydrophobic surfaces. Furthermore, by varying the viscosity ratio and the inclination angle of the plate, we found that the droplet's migration speed decreases with an increase in the viscosity ratio. In particular, two vortices appear inside the droplet at a high viscosity ratio, whereas only one vortex is present at a low viscosity ratio.
\end{abstract}
	
\end{frontmatter}

%\linenumbers

\section{Introduction}
Thermocapillary flow is a fluid motion phenomenon caused by interfacial tension imbalance due to temperature or composition differences. Physically, for most fluids, surface tension changes with temperature, which means that when an immiscible droplet or bubble suspended in a liquid is exposed to unequal temperature environments, this imbalance in tension drives the droplet or bubble to move. This characteristic makes thermocapillary migration of droplets or bubbles significant in microfluidics \cite{Haeberle_LoC2007,Nguyen_Applphy2005}, material synthesis \cite{Serra_Langmuir2007,Carroll_Langmuir2008,Chine_ICHMT2006}, the design and operation of integrated DNA analysis devices \cite{Burns_NAS1996,Gallardo_science1999}, and disease diagnosis \cite{Garcia-Cordero_labonachip2017}. Therefore, understanding the thermocapillary migration of droplets has profound implications for these areas.

Research on the thermocapillary migration of droplets can be traced back to Bouasse’s experiments \cite{Bouasse_Delagrave1924}, which discovered that the presence of a temperature gradient tends to induce the droplet to climb upward against gravity. Subsequently, Young et al. \cite{Young_JFM1959} experimentally demonstrated that Marangoni stess induced by temperature could cause small bubble in the fuild to move downward against bouyancy. Since then, numerous scholars have endeavored to uncover the mechanisms of thermocapillary flows through theoretical and experimental studies. Brzoska et al. \cite{Brzoska_Langmuir1993} experimentally investigated the motion of droplets on non-wettable surface under a horizontal temperature gradient, and found that as the temperature gradient increased, the droplet exhibited three behaviors: stationary, migratory, and stretching. Similarly, Brochard \cite{Brochard_Langmuir1989} theoretically studied the motion of droplets on substrate with chemical or temperature gradients, and observed that droplet moves towards regions of higher surface energy. Combined with Brochard’s work \cite{Brochard_Langmuir1989}, Ford and Nadim \cite{Ford_POF1994} used lubrication theory to study the migration velocity of two-dimensional arbitrarily shaped droplets on a substrate with a temperature gradient. Furthermore, Chen et al. \cite{Chen_jap2005} validated Ford and Nadim’s theory though experiment. Builded upon the research of Ford and Nadim \cite{Ford_POF1994}, Pratap et al. \cite{Pratap_langmuir2008} conducted similar experiments and obtained numerical solutions for the motion of three-dimensional droplet on substrate with temperature gradients. Apart from these experimental and theoretical studies, the rapid development of computer technology in recent years has facilitated the use of numerical simulations to investigate thermocapillary flows \cite{Le_ATE2017,Qiao_ATE2018,Huang_ATE2018}. For instance, Sui \cite{Sui_POF2014} employed the level-set method to study the thermocapillary migration of droplets, and results show that changes in viscosity and contact angle could cause droplet to migrate towards the heated area. Based on this work, Fath and Bothe \cite{Fath_IJMF2015} adopted the volume of fluid method to conduct numerical studies of the thermocapillary migration of a three-dimensional droplet, and the droplet dynamics are consistent with Sui's research \cite{Sui_POF2014}. Simultaneously, Karapetsas et al. \cite{Karapetsas_Langmuir2013} numerically simulated the thermocapillary migration of droplets on inclined surfaces, which found that thermocapillary effects could enhance the spreading rate and "stick-slip" behavior of sessile droplets on inclined substrate. More recently, Nguyen et al. \cite{Nguyen_JFE2022} investigated the thermocapillary migration of compound droplets in narrow channels, and observed that a process of deceleration followed by acceleration as the droplets moved through the channels.

Although these studies have provided a fundamental understanding of thermocapillary motion of droplets, they have focused on normal fluids such as water and air, where surface tension monotonically decreases with increasing temperature. However, experiments have shown that dilute aqueous solutions of long-chain alcohols exhibit a parabolic relationship between surface tension and temperature, with a defined minimum. Such fluids, including butanol and pentanol, are referred to as self-rewetting fluids \cite{Vochten_jcis1973,Petre_JCIS1984}. Recent experiments indicated that these fluids have significant application potential in droplet manipulation \cite{Wu_IJTS2015,Singh_TSEP2021}. In recent years, some scholars have conducted research on self-rewetting fluids. Savino et al. \cite{Savino_IJHMT2017} experimentally and theoretically investigated the two-phase flow, heat transfer, and drying behavior of self-rewetting fluids in a grooved heat pipe model. Balla et al. \cite{Balla_JFM2019} numerically studied the migration behavior of bubbles in non-isothermal self-rewetting fluids. Elbousefi et al. \cite{Elbousefi_IJHMT2023} focused on the thermocapillary convection in superimposed layers of self-rewetting fluids. However, these studies did not consider fluid-surface interactions. To adress this issue, Chaudhury et al. \cite{Chaudhury_Langmuir2015} theoretically constructed the spreading behavior of self-rewetting droplets on non-isothermal surfaces and used the volume of fluid method to validate his theory through numerical simulations. Laterly, Xu et al. \cite{Xu_JFM2021} extended this work to the three-dimensional spreading and movement of droplets using the conservative level-set method. Despite numerous experimental and numerical studies have been conducted on the thermocapillary migration of self-rewetting droplets, these studies have predominantly focused on horizontal surfaces. In practical applications, it is not always feasible to maintain the working surface in a horizontal state \cite{Abe_AAS2006}. Therefore, investigating the migration of droplets on inclined surfaces is essential. Mamalis et al's \cite{Mamalis_APL2016} experiments on self-rewetting droplets on uniformly heated inclined plates discovered that the self-rewetting droplets can also move upwards along the inclined plate. Based on this works, Ye et al. \cite{Ye_POF2018} considered the influencse of the inclination angle, Bond number, and capillary number and discovered the droplet exhibited complex deformations. Note that these available works are confined to the uniformly inclined substrate, to the best of our knowledge, no researchers have investigated the motion of droplets on inclined surfaces with temperature gradients.

Framed in this general background, the current work aims to numerically study the thermocapillary migration of self-rewetting droplets on inclined surfaces. As far as the numerical method is concerned, the lattice Boltzmann (LB) method is adopted in this work. Unlike the conventional computational fluid dynamics methods mentioned above, the LB method features a simple algorithm and inherent parallelism, and it can handle complex boundaries \cite{Li_PECS2016,Huang_CF2022}, making it a popular method for simulating complex flows \cite{Kruger_SIP2017}, including thermocapillary flows \cite{Elbousefi_IJHMT2023,Liu_jcp2014,Wang_PRE2023}. The remainder of this paper is organized as follows: The phase field method is presented in the next section (Section 2). Section 3 introduces the LB method for the temperature field. In Section 4, we provided a detailed description of the problem. In Section 5, the employed LB model was validated, and the results and discussion regarding the problem were presented. Finally, Section 6 provides brief conclusions.

\section{Governing equations} 
The phase field method is one of the most commonly used methods for simulating multiphase flows \cite{Badalassi_JCP2003,Chiu_JCP2011,Hua_JCP2011}. It belongs to the class of interface capturing methods, characterized by describing the phase interface with a rapidly varying but continuous phase field variable, also known as the order parameter. In traditional phase field models, the most frequently used interface capturing equation is the Cahn-Hilliard equation \cite{Chiu_JCP2011,Hua_JCP2011}. However, due to the non-local collision process of the LB model corresponding to the Cahn-Hilliard equation \cite{Liang_PRE2018}, this paper adopts the following second-order conservative Allen-Cahn equation \cite{Wang_Capillarity2019}

\begin{equation}
	\frac{{\partial \phi }}{{\partial t}} + \nabla  \cdot \left( {\phi {\bf{u}}} \right) = \nabla  \cdot {M_\phi }\left( {\nabla \phi  - \frac{{\nabla \phi }}{{\left| {\nabla \phi } \right|}}\frac{{1 - {\phi ^2}}}{{\sqrt 2 W}}} \right),
\end{equation}
where $\phi $ is the order parameter, ${M_\phi }$ represents the mobility, and $W$ is the interface thickness. In this work, we use ${{\phi _{A}} = 1}$ to stand for the droplet and ${{\phi _{B}} = -1}$ to represent the external fluid. Thus, the interface between these two fluids can be described by $\phi  = 0$.

In addition to the aforementioned interface capturing equation, we also need the governing equations to describe the two-phase flow. Assuming the two fluids are immiscible and incompressible, the Navier-Stokes equations can be written as \cite{Unverdi_JCP1992}
\begin{equation}
	\begin{array}{c}
		\nabla  \cdot {\bf{u}} = 0,\\
		\rho \left( {\frac{{\partial {\bf{u}}}}{{\partial t}} + {\bf{u}} \cdot \nabla {\bf{u}}} \right) =  - \nabla p + \nabla  \cdot \left[ {\mu \left( {\nabla {\bf{u}} + \nabla {{\bf{u}}^{\bf{T}}}} \right)} \right] + {{\bf{F}}_s} + {\bf{G}},
	\end{array}
\end{equation}
where $\rho $ is the fluid density, and $p, {\bf{u}}, \mu$ represent the hydridynamic pressure, fluid velocity and dynamic viscosity, respectively. ${\bf{G}}$ represents the body force, and ${{\bf{F}}_s}$ is the surface tension force. which can be written as \cite{Wang_Capillarity2019}
\begin{equation}
	{{\bf{F}}_s} = \frac{{3\sqrt 2 }}{4}\left[ {{{\left| {\nabla \phi } \right|}^2}\nabla \sigma  - \nabla \sigma  \cdot \left( {\nabla \phi \nabla \phi } \right) + \frac{\sigma }{{{W^2}}}{\mu _\phi }\nabla \phi } \right],
\end{equation}
where ${{\mu _\phi }}$ is the chemical potential, defined as \cite{Wang_Capillarity2019}
\begin{equation}
	{\mu _\phi } = \left( {\phi  - {\phi _{A}}} \right)\left( {\phi  - {\phi _{B}}} \right)\left[ {\phi  - 0.5\left( {{\phi _{A}} + {\phi _{B}}} \right)} \right] - {W^2}{\nabla ^2}\phi.
\end{equation}
Note that the current work focuses on self-rewetting fluids, and thus the surface tension $\sigma $ and temperature $T$ have a quadratic relationship, which can be given by \cite{Xu_JFM2021}
\begin{equation}
	\sigma  = {\sigma _{ref}} - {\sigma _T}\left( {T - {T_c}} \right) + {\sigma _{TT}}{\left( {T - {T_c}} \right)^2},
\end{equation}
where ${\sigma _{ref}}$ is the surface tension at the reference temperature $T_c$, ${\sigma _T}{\left. { = \frac{{d\sigma }}{{dT}}} \right|_{{T_c}}}$ and ${\sigma _{TT}}{\left. { = \frac{{{d^2}\sigma }}{{\partial {T^2}}}} \right|_{{T_c}}}$ are the coefficient of the surface tension with respect to temperature. Apparently, the value of $\sigma_{TT}$ is set to 0, the corresponding fluid is the normal fluid, which surface tension decreases linearly with temperature.

In thermocapillary flow, obtaining the time evolution of the temperature field is crucial for understanding thermal multiphase flow. By neglecting the effects of viscous heat dissipation. The governing equation for the temperature field can be given as follows \cite{Wang_PRE2023}
\begin{equation}
	\rho c_p\left(\frac{\partial T}{\partial t}+\mathbf{u} \cdot \nabla T\right)=\nabla \cdot \lambda \nabla T,
\end{equation}
where $\lambda $ is the thermal conductivity, and ${c_p}$ is the specific-heat capacity.

\section{Methodology} 
\subsection{LB method for two-phase thermocapillary flows}
According to the collision operator used, the LB method can be divided into three type, the single-relaxation-time or so-called Bhatnagar-Gross-Krook (BGK) method \cite{Qian_EL1992}, the two-relaxation-time method \cite{Ginzburg I_CCP2008}, and the multiple-relaxation-time (MRT) method \cite{Lallemand_PRE2000}. In this work, we use the BGK model because of its computational simplicity and efficiency. For two-phase flow, the evolution equations for all three fields adopt the standard two-dimensional nine-velocity (D2Q9) lattice model, which is defined as
\begin{equation}
	\mathbf{c}_i= \begin{cases}c(0,0)^{\mathrm{T}}, & i=0, \\ c(\cos [(i-1) \pi / 2], \sin [(i-1) \pi / 2])^{\mathrm{T}}, & i=1,2,3,4, \\ \sqrt{2}(\cos [(2 i-1) \pi / 4], \sin [(2 i-1) \pi / 4])^{\mathrm{T}}, & i=5,6,7,8,\end{cases}
\end{equation}
where $c=\Delta x / \Delta t$ is the lattice speed. For the D2Q9 lattice, the sound speed $c_s$ is given as $c_s=c / \sqrt{3}$, and the weight coeficient is expressed as $\omega_0=4 / 9, \omega_{1,2,3,4}=4 / 9$, and $\omega_{5,6,7,8}=1 / 36$. 

To simulate the thermocapillary flows with the LB method, we adopt three LB equations, one of which is used to solve the Allen-Cahn equation, the others are utilized to solve the incompressible Navier-Stokes equations and the temperture field, respectively \cite{Wang_PRE2016}. The LB equation for Allen-Cahn equation can be written as
\begin{equation}
	{g_i}\left( {{\bf{x}} + {{\bf{c}}_i}\Delta t,t + \Delta t} \right) = {g_i}\left( {{\bf{x}},t} \right) - \frac{1}{{{\tau _\phi }}}\left[ {{g_i}\left( {{\bf{x}},t} \right) - g_i^{eq}\left( {{\bf{x}},t} \right)} \right] + \Delta t\left( {1 - \frac{1}{{2{\tau _\phi }}}} \right){G_i}\left( {{\bf{x}},t} \right),
\end{equation}
where ${g_i}\left( {{\bf{x}},t} \right)$ is the distribution function of the order parameter with respect to the position ${\bf{x}}$ and the time $t$, and ${{\tau _\phi }}$ is the relaxation time of phase field, which can be written as ${M_\phi } = c_s^2\left( {{\tau _\phi } - 0.5} \right)\Delta t$, $g_i^{eq}$ and ${G_i}$ represent the equilibrium distribution function and source term, respectively, and they are defined as
\begin{equation}
	g_i^{e q}(\mathbf{x}, t)=\omega_i \phi\left(1+\frac{\mathbf{c}_i \cdot \mathbf{u}}{c_s^2}\right),
\end{equation}
\begin{equation}
	G_i(\mathbf{x}, t)=\omega_i \frac{\mathbf{c}_i \cdot\left[\partial_t(\phi \mathbf{u})+c_s^2 \frac{\nabla \phi}{\nabla \phi \mid} \frac{1-\phi^2}{\sqrt{2} W}\right]}{c_s^2},
\end{equation}
the order parameter $\phi $ can be computerd from ${g_i}$ \cite{Wang_PRE2016}
\begin{equation}
	\phi(\mathbf{x}, t)=\sum_i g_i(\mathbf{x}, t).
\end{equation}

\par In addition to the LB equation for the Allen-Cahn equation, we also need another LB model to solve the Navier-Stokes equations. Based on previous work, we can use the following LB equation to solve the Navier-Stokes equations \cite{Liang_PRE2018}
\begin{equation}
	f_i\left(\mathbf{x}+\mathbf{c}_i \Delta t, t+\Delta t\right)=f_i(\mathbf{x}, t)-\frac{1}{\tau_f}\left[f_i(\mathbf{x}, t)-f_i^{e q}(\mathbf{x}, t)\right]+\Delta t\left(1-\frac{1}{2 \tau_f}\right) F_i(\mathbf{x}, t),
\end{equation}
where ${f_i}$ is the discrete velocity distribution function, the equilibrium distribution function $f_i^{eq}$ is defined as \cite{Liang_PRE2018}
\begin{equation}
	f_i^{e q}=\left\{\begin{array}{l}
		\frac{p}{c_x^2}\left(w_0-1\right)+\rho s_i(\mathbf{u}), i=0 \\
		\frac{p}{c_x^2} w_i+\rho s_i(\mathbf{u}), i \neq 0
	\end{array}\right.,
\end{equation}
with
\begin{equation}
	s_i(\mathbf{u})=\omega_i\left[\frac{\mathbf{c}_i \cdot \mathbf{u}}{c_s^2}+\frac{\left(\mathbf{c}_i \cdot \mathbf{u}\right)^2}{2 c_s^4}-\frac{\mathbf{u} \cdot \mathbf{u}}{2 c_s^2}\right],
\end{equation}
the discrete forcing term ${F_i}$ is designed as
\begin{equation}
	F_i=\omega_i\left[\mathbf{u} \cdot \nabla \rho+\frac{\mathbf{c}_i \cdot\left(\mathbf{F}_s+\mathbf{G}\right)}{c_s^2}+\frac{\mathbf{u} \nabla \rho:\left(\mathbf{c}_i \mathbf{c}_i-c_s^2 \mathbf{I}\right)}{c_s^2}\right].
\end{equation}
Through Chapman-Enskog analysis, the incompressible Navier-Stokes equations are recovered to second-order accuracy, yielding the relaxation time for the velocity field as \cite{Liang_PRE2018}
\begin{equation}
	\tau_f=\frac{1}{2}+\frac{\mu}{\rho c_s^2 \Delta t},
\end{equation}
and the macroscopic quantities can be computed from \cite{Liang_PRE2018}
\begin{equation}
	\begin{aligned}
		\rho \mathbf{u} & =\sum_i \mathbf{c}_i f_i+0.5 \Delta t\left(\mathbf{F}_s+\mathbf{G}\right), \\
		p & =\frac{c_i^2}{\left(1-\omega_n\right)}\left[\sum_{i \neq 0} f_i+\frac{\Delta t}{2} \mathbf{u} \cdot \nabla \rho+\rho s_0(\mathbf{u})\right].
	\end{aligned}
\end{equation}
\par For the temperature field, the improved LB model proposed by Wang et al \cite{Wang_PRE2023} is adopted in the current work, and the corresponding LB equation is express as
\begin{equation}
	\rho c_p h_i\left(\mathbf{x}+c_i \Delta t, t+\Delta t\right)-h_i(\mathbf{x}, t)=\left(\rho c_p-1\right) h_i\left(\mathbf{x}+c_i \Delta t, t\right)-\frac{1}{\tau_h}\left[h_i(\mathbf{x}, t)-h_i^{e q}(\mathbf{x}, t)\right]+\Delta t \hat{F}_i(\mathbf{x}, t),
\end{equation}
where ${h_i}$ is the temperature distribution function, ${\tau _h}$ is the dimentionless relaxation time defined to ${\lambda _h} = 0.5 + {\lambda  \mathord{\left/
		{\vphantom {\lambda  {c_s^2}}} \right.
		\kern-\nulldelimiterspace} {c_s^2}}\Delta t$, In addition, the equilibrium functions $h_i^{eq}$ is given by \cite{Wang_PRE2023}
\begin{equation}
	h_i^{e q}(\mathbf{x}, t)=\omega_i T \text {, }
\end{equation}
and the discrete source term is $\hat{F}_i(\mathbf{x}, t)$ is determined by \cite{Wang_PRE2023}
\begin{equation}
	{\widehat F_i}\left( {{\bf{x}},t} \right) =  - {\omega _i}\rho {c_p}{\bf{u}} \cdot \nabla T.
\end{equation}
Further, the temperature can be computerd as \cite{Wang_PRE2023}
\begin{equation}
	T=\sum_i h_i .
\end{equation}

\subsection{Wetting boundary condition}
Due to the consideration of surface wettability effects during the simulation, it is essential to construct a wetting boundary condition that accounts for the contact angle between the phase interface and the solid surface. To this end, we employ a geometric formulation-based wetting boundary condition \cite{Ding_PRE2007}, which is defined as
\begin{equation}
	\mathbf{n}_w \cdot \nabla \phi=-\tan \left(\frac{\pi}{2}-\theta\right)\left|\mathbf{n}_\tau \cdot \nabla \phi\right|,
\end{equation}
where $\mathbf{n}_w$ is the unit vector normal to the solid surface, and $\mathbf{n}_\tau$ is the unit vector tangential to the solid surface. With the second-order diffence schme, the wetting boundary condition can be rewritten as
\begin{equation}
	\frac{\phi_{x, 1}-\phi_{x, 0}}{\delta_x}=-\tan \left(\frac{\pi}{2}-\theta\right)\left|\mathbf{n}_\tau \cdot \nabla \phi\right|,
\end{equation}
here $\phi_{x, 0}$ represents the order parameter beneath the first layer, which can be determined by $\phi_{x, 0}=2 \phi_{x, 1 / 2}-\phi_{x, 1}$ \cite{Liang_PRE2019}. In addition, the right side of Eq. (24) can be determined by using a second-order extrapolation scheme
\begin{equation}
	\mathbf{n}_\tau \cdot \nabla \phi=\frac{\partial \phi_{x, 1 / 2}}{\partial x}=1.5 \frac{\partial \phi_{x, 1}}{\partial x}-0.5 \frac{\partial \phi_{x, 2}}{\partial x},
\end{equation}
In such a case, the wetting boundary condition could easily realized by using Eq. (23) and Eq. (24).

\section{Problem statement}
\begin{figure}[H]
	\centering
	\includegraphics[width=0.6\textwidth]{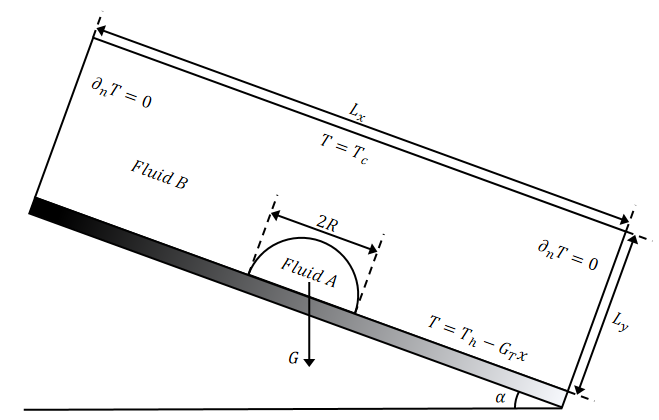}
	\caption{Schematic of a droplet on an inclined solid substrate with a temperature gradient, where $T = {T_c}$ and $T = {T_h} - {G_T}x$ are the temperature of the top wall and the bottom wall, respectively, and the other walls are adidabatic.}
	\label{fig5}
\end{figure}
Fig. 1 shows the thermocapillary migration of a droplet on an inclined surface. Initially, a hemispherical droplet rests in the middle of a substrate with a temperature gradient applied to the solid surface. The droplet radius is set to be $R = 50\Delta x$, and the computational rectangular domain is ${L_x} \times {L_y} = 16R \times 3R$. No-slip condidtions are applied to all boudaries of the computational domain, which are considered as stationary walls. For the temperature field, the left and right boundaries are adiabatic, i.e., ${\partial _n}T = 0$. The top boundary is set to a constant low temperature which can be considered the ambient temperature, and the bottom boundary has a linear temperature gradient, i.e., $T\left( {x,0} \right) = {T_H} - {G_T}x$. Following the previous work \cite{Liu_jcp2015}, the high temperature ${T_h}$ and the low temperature ${T_c}$ are set to $80$ and $0$, respectively.

The thermocapillary migration of self-rewetting droplet can be characterizes by using the following dimensionless parameters: Reynolds number (Re), capillary number (Ca), Marangoni number (Ma) and the Bond number (Bo), and they are defined as
\begin{equation}
	{\mathop{\rm Re}\nolimits}  = \frac{{{\rho _A}UR}}{{{\mu _A}}},Ca = \frac{{U{\mu _A}}}{{{\sigma _{ref}}}},Ma = \frac{{{\rho _A}{c_{pA}}}}{{{\lambda _A}}},Bo = \frac{{{\rho _A}g{R^2}}}{{{\sigma _{ref}}}},
\end{equation}
where $U =  - \frac{{{\sigma _T}{G_T}R}}{{{\mu _A}}}$ represents the system’s characteristic velocity, g is the gravity, and A and B indicate the two different fluids involved. 

\section{Result and Discussion}
In this section, we intend to simulate the thermocapillary migration of a self-rewetting droplet on an inclined surface, focusing on the detaile investigation of the Marangoni number, surface wettability, and the viscosity ratio. In this simulation, the dimensionless parameters used are generally consisitant with those in Liu et al.'s study \cite{Liu_jcp2015}
\begin{equation}
	{\mathop{\rm Re}\nolimits}  = 10,Ca = 4.4 \times {10^{ - 2}},\frac{{{\nu _A}}}{{{\nu _B}}} = 1,\frac{{{\lambda _A}}}{{{\lambda _B}}} = 1.
\end{equation}
In addition, since the effect of gravity is considered in this model, the Bond number is set to $0.2$, which is frequently used in previous stuies \cite{Balla_JFM2019,Ye_POF2018}. Based on experimental research on self-rewetting fluids, other parameters are set as ${T_m} = 20,{{{c_{pB}}} \mathord{\left/
		{\vphantom {{{c_{pB}}} {{c_{pA}} = 0.2}}} \right.
		\kern-\nulldelimiterspace} {{c_{pA}} = 0.2}},{{{\rho _B}} \mathord{\left/
		{\vphantom {{{\rho _B}} {{\rho _A} = 0.8}}} \right.
		\kern-\nulldelimiterspace} {{\rho _A} = 0.8}}$ \cite{Vochten_jcis1973}.The fluid density distribution in a two-phase system physically aligns with the order parameter. To maintain this physical property, the fluid density, dynamic viscosity, thermal conductivity and specific heat capacity should be represented using linear interpolation.
\begin{equation}
	\begin{aligned}
		& \rho=\frac{\rho_A-\rho_B}{\phi_A-\phi_B}\left(\phi-\phi_B\right)+\rho_B, \mu=\frac{\mu_A-\mu_B}{\phi_A-\phi_B}\left(\phi-\phi_B\right)+\mu_B, \\
		& \lambda=\frac{\lambda_A-\lambda_B}{\phi_A-\phi_B}\left(\phi-\phi_B\right)+\lambda_B, c_p=\frac{c_{p A}-c_{p B}}{\phi_A-\phi_B}\left(\phi-\phi_B\right)+c_{p B},
	\end{aligned}
\end{equation}

\subsection{Numerical Validation}
In Ref. \cite{Wang_PRE2023}, the current phase-field LB model has been confirmed to accurately simulate thermocapillary flows without fluid-surface interactions. To further validate the current method, we simulate the thermocapillary migration of normal droplet on the solid surface by comparing with Liu et al's work \cite{Liu_jcp2015}. In this simulation, the parameters are set as $Re=10$, $Ca=4.4 \times 10^{-2}$, $Ma=10$.

Fig. 2 shows the streamline plot around the droplet (left) and the isotherm plot (right) at a contact angle of $60^{\circ}$ and the time step of $t=2 \times 10^5$. From this figure, we can observe that our results align well with those of Liu et al. \cite{Liu_jcp2015}. It is interesting to note that an apparent vortex is present both inside and outside the droplet, with the internal vortex located on the left side of the droplet and the external vortex on the upper side of the droplet.
\begin{figure}[H]
	\centering
	\includegraphics[width=0.8\textwidth]{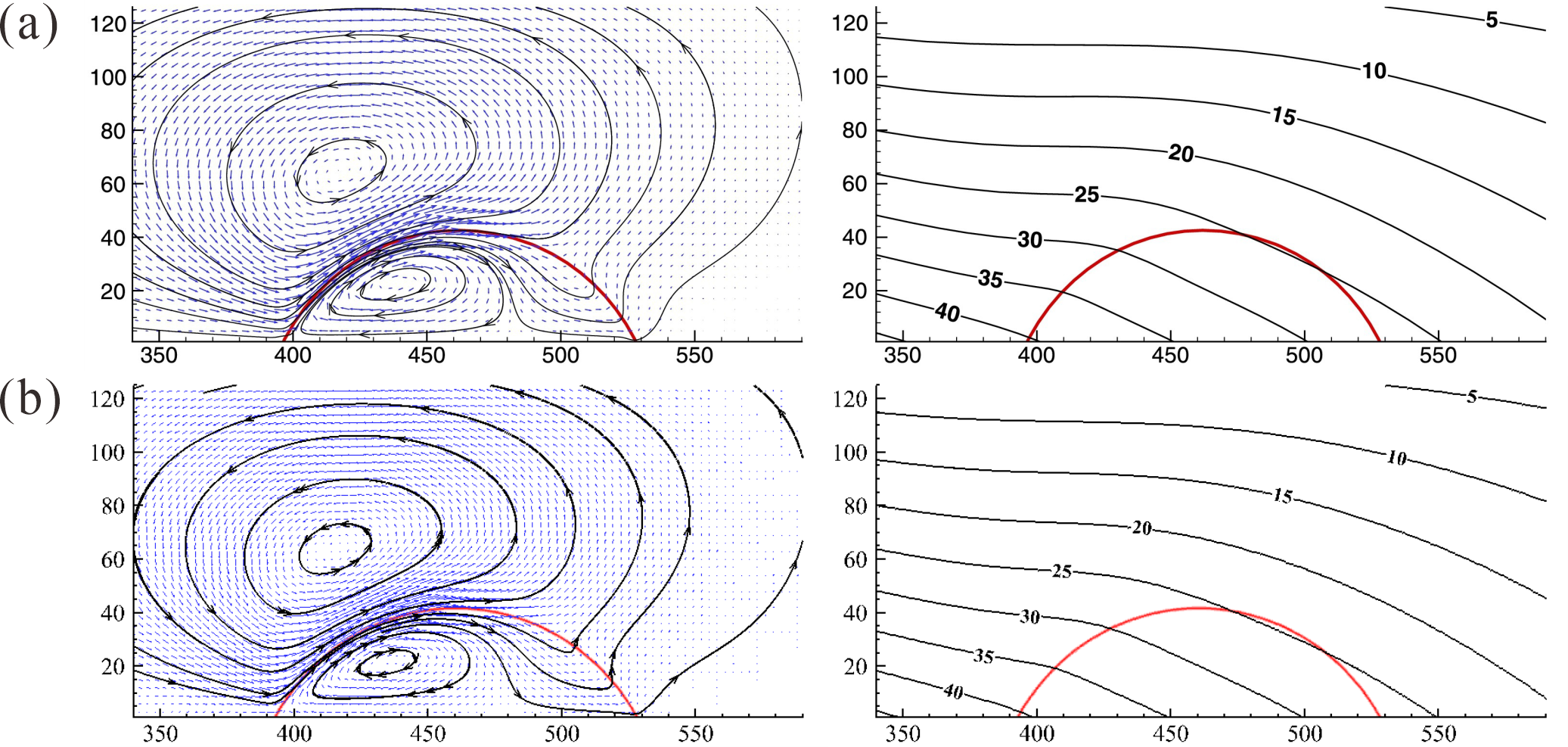}
	\caption{The streamline (the left plane) and the temperture (the right plane) surrounding the moving droplet at the contact angle $\theta = 60^{\circ}$, (a) the numerical results obtained by Liu et al. \cite{Liu_jcp2015}, (b) the present date, in which the red lines are $\phi = 0$, the blue lines with arrows are the velocity vectors, and the black lines with arrows are the streamlines}.
	\label{fig2}
\end{figure}
\noindent
In addition to the qualitative study, we alse compared the relationship between the centroid of the droplet and time during migration under different contact angles. and its definition can be expressed as
\begin{equation}
	{x_s}\left( t \right) = \frac{{\sum\nolimits_{\bf{x}} {x({\bf{x}},t){\rho _A}\left( {{\bf{x}},t} \right)} }}{{\sum\nolimits_{\bf{x}} {{\rho _A}\left( {{\bf{x}},t} \right)} }}.
\end{equation}
As shown in Fig. 3, one can cleary find that our results fit well with those from previous studies across various contact angles, thus validating the LB model's effectivenless in handing fluid-surface interactions.
\begin{figure}[H]
	\centering
	\includegraphics[width=0.6\textwidth]{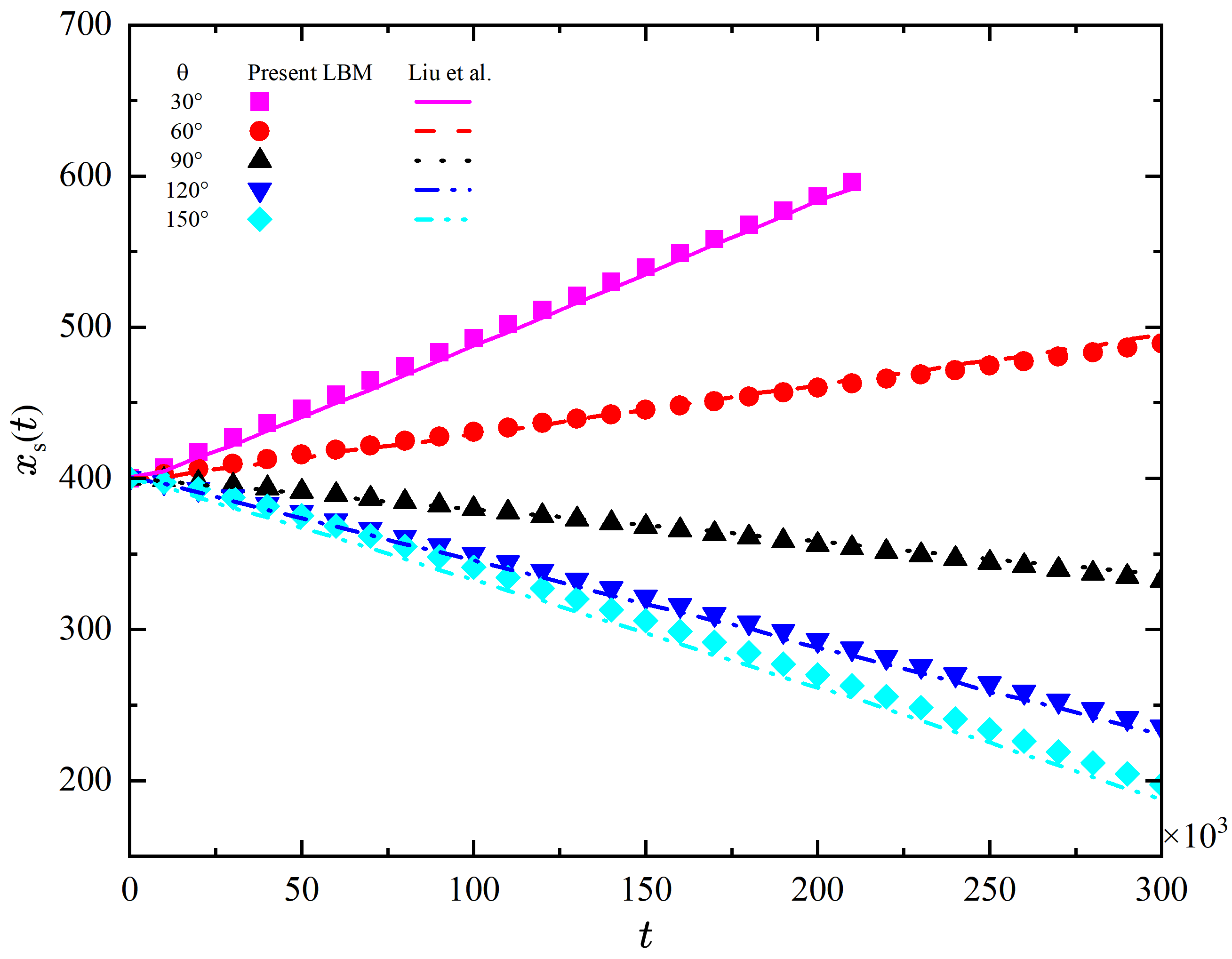}
	\caption{Time evolution of the $x_s$ at various contact angles.}
	\label{fig3}
\end{figure}
	
\subsection{The impact of Marangoni number on droplet migration}	
Investigation the influence of marangoni number on droplet migration will be explored in this subsection, and the value of Marangoni number is set to be $50$, $100$, $150$, and $200$, respectively. Fig. 4 illustrates the comparison of droplet morphologies at different Marangoni numbers under the same Fourier number (${F_o}$) which is defined as ${F_o} = {{\lambda t} \mathord{\left/
		{\vphantom {{\lambda t} {\rho {c_p}L_y^2}}} \right.
		\kern-\nulldelimiterspace} {\rho {c_p}L_y^2}}$. It is observed that the droplet exhibits overall symmetry, and with an increase in Marangoni number, the length of the droplet's contact line elongates, indicating a more pronounced droplet extension.
\begin{figure}[H]
	\centering
	\includegraphics[width=0.8\textwidth]{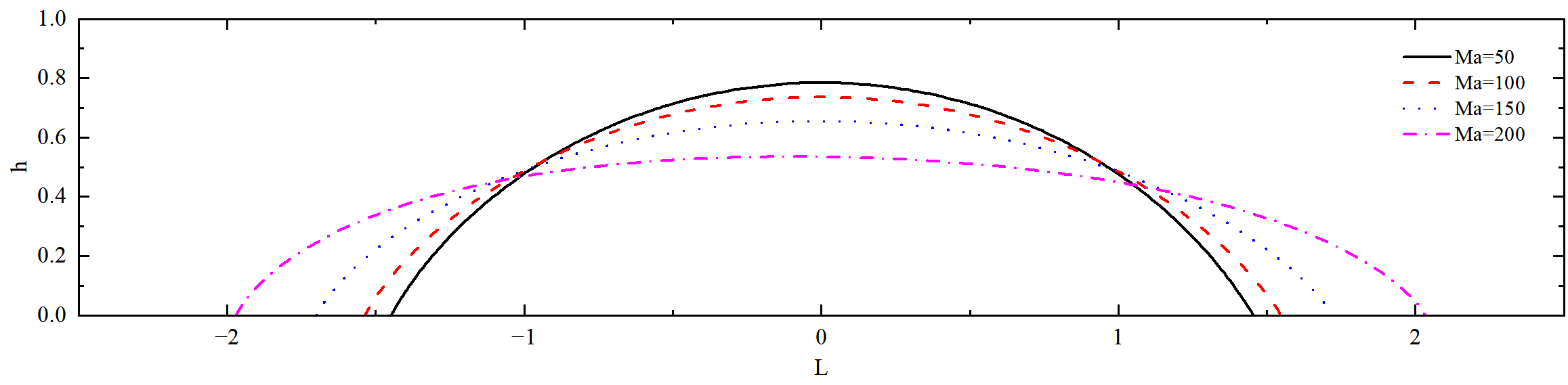}
	\caption{Contour plot of droplets at different Marangoni number, for ${F_o} = 0.44$, $\theta  = 60^\circ $.}
	\label{fig5-1}
\end{figure}
\begin{figure}[H]
	\centering
	\includegraphics[width=0.8\textwidth]{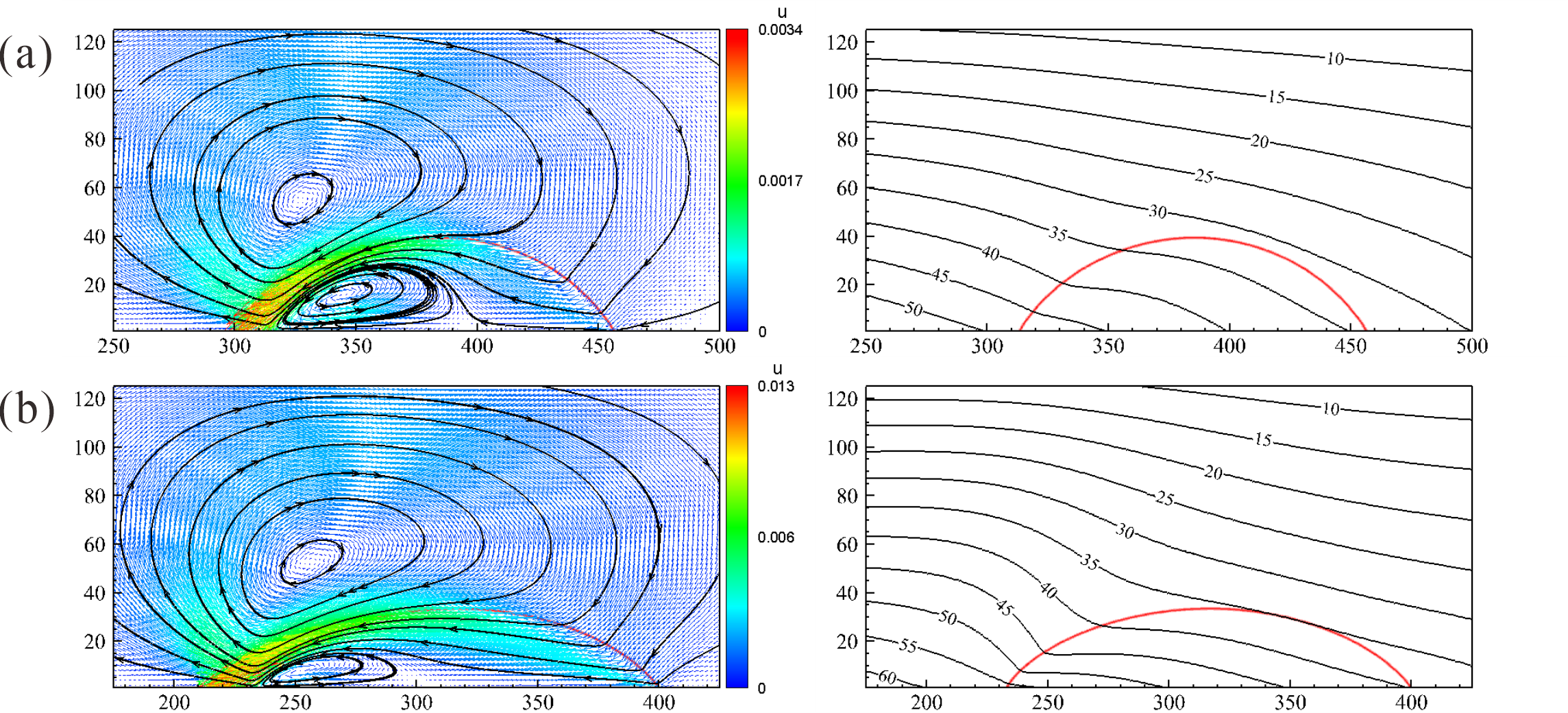}
	\caption{The flow field (the left plane) and the temperature field (the right plane) surrouding the moving droplet at the contact angle $\theta  = 60^\circ $, (a) Ma = 50, (b) Ma = 150. The red lines are $\phi  = 0$, the colorful lines with arrows are the velocity vectors, and the black lines with arrows are the streamlines.}
	\label{fig5-5}
\end{figure}
To investigate the mechanism behind the droplet deformation for different Marangoni numbers, we also conducted a force analysis on the droplet, which builds upon the work of Xu et al \cite{Xu_JFM2021}. The schematic diagram of the force analysis is shown in Fig. 6, in which the ${F_{s1}}$ represents the surface tension exerted on the bottom of the droplet along the horizontal direction to the right, ${F_{s2}}$ represents the surface tension exerted on the droplet along the tangent direction to the interface on the symmetrical plane, ${F_p}$ denotes the resultant pressure force exerted on the droplet's symmetrical plane, obtained by integrating the internal pressure $p$ along the symmetry line, and ${F_\mu }$ signifies the viscous force, which is formed at the bottom due to the internal vortex within the droplet. The internal flow induced by the marangoni effect and moving in a clockwise direction causes the viscous force to be directed towards the contact line of the droplet. The resultant force acting on this control force can be determined as
\begin{equation}
	{F_e} = {F_{s1}} - {F_{s2}}\cos \theta  - {F_\mu } - {F_p},
\end{equation}
\begin{figure}[H]
	\centering
	\includegraphics[width=0.6\textwidth]{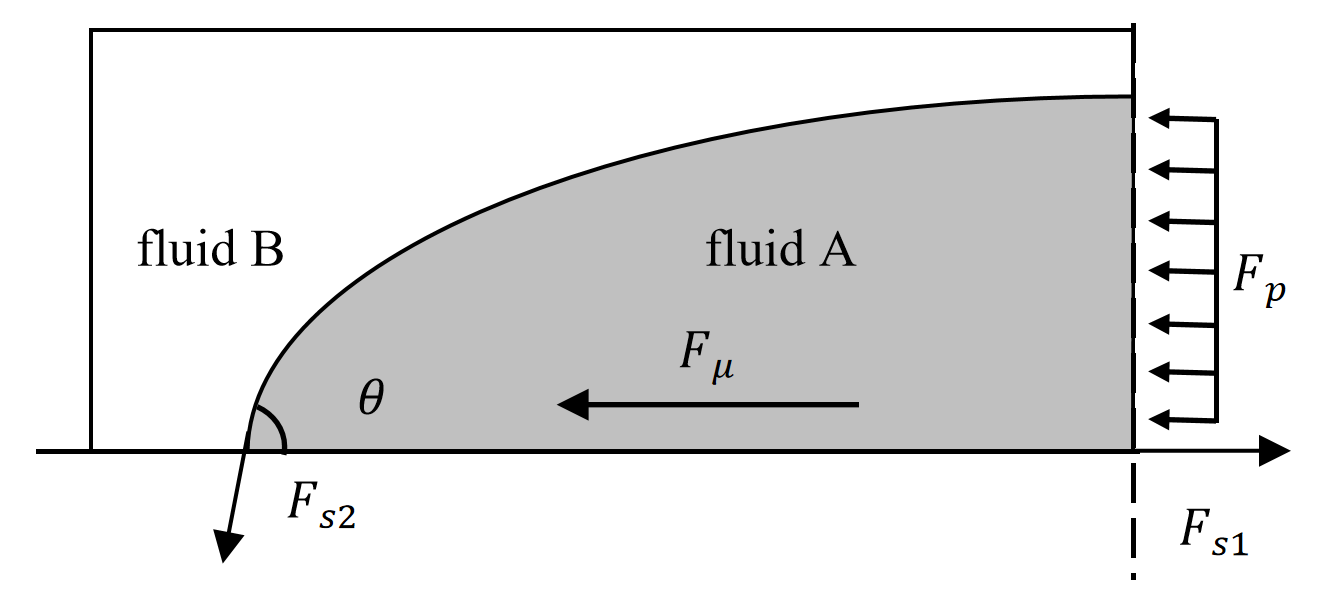}
	\caption{Schematic diagram of the theoretical force model on a droplet, in which ${F_{s1}}$ represents the surface tension at the symmetric point, ${F_{s2}}$ denotes the surface tension at the contact line, ${F_\mu }$ indicates the viscous drag on the droplet, ${F_p}$ stands for the integral of the internal pressure along the droplet's symmetry line, and $\theta $ is the contact angle of the droplet.}
	\label{fig5-2}
\end{figure}
\noindent
where $\theta $ represents the equilibrium contact angle of the droplet. When the resultant force ${F_e} < 0$, the droplet tends to spread due to excessive tension, while when ${F_e} > 0$, the droplet remains in equilibrium. In this context, expressions for each force are calculated as follows
\begin{equation}
	\left\{ \begin{array}{l}
		{F_{s2}} = {\left. {\sigma \left( {x,{M_1}} \right)} \right|_{x = 0}}\\
		{F_{s2}} = {\left. {\sigma \left( {x,{M_1}} \right)} \right|_{x =  - R}}
	\end{array} \right.,\left\{ \begin{array}{l}
		{F_\mu } = \mu \frac{{\partial u}}{{\partial y}} \cdot R\\
		{F_p} = \frac{{\int {{{\left. {\sigma \left( {x,{M_1}} \right)} \right|}_{x = 0}}} }}{R}
	\end{array} \right..
\end{equation}
We track the resultant force ${F_e}$ experienced by the droplet at different Marangoni numbers as a function of the ${F_o}$, as shown in Fig. 7. Under the considered Marangoni numbers, ${F_e}$ is consistently negative, and it decreases with an increase in Marangoni number. Interestingly, when the Marangoni number is 50, the value of ${F_e}$ remains nearly constant, but as the Marangoni number increases, ${F_e}$ decreases gradually with the evolution of ${F_o}$. This can be attributed to the droplet gradually migrating towards the hotter end during motion, resulting in an increased tension differential on both sides of the droplet due to the nonlinear variation of surface tension with temperature. Consequently, maintaining the droplet's original shape during migration becomes increasingly difficult.
\begin{figure}[H]
	\centering
	\includegraphics[width=0.6\textwidth]{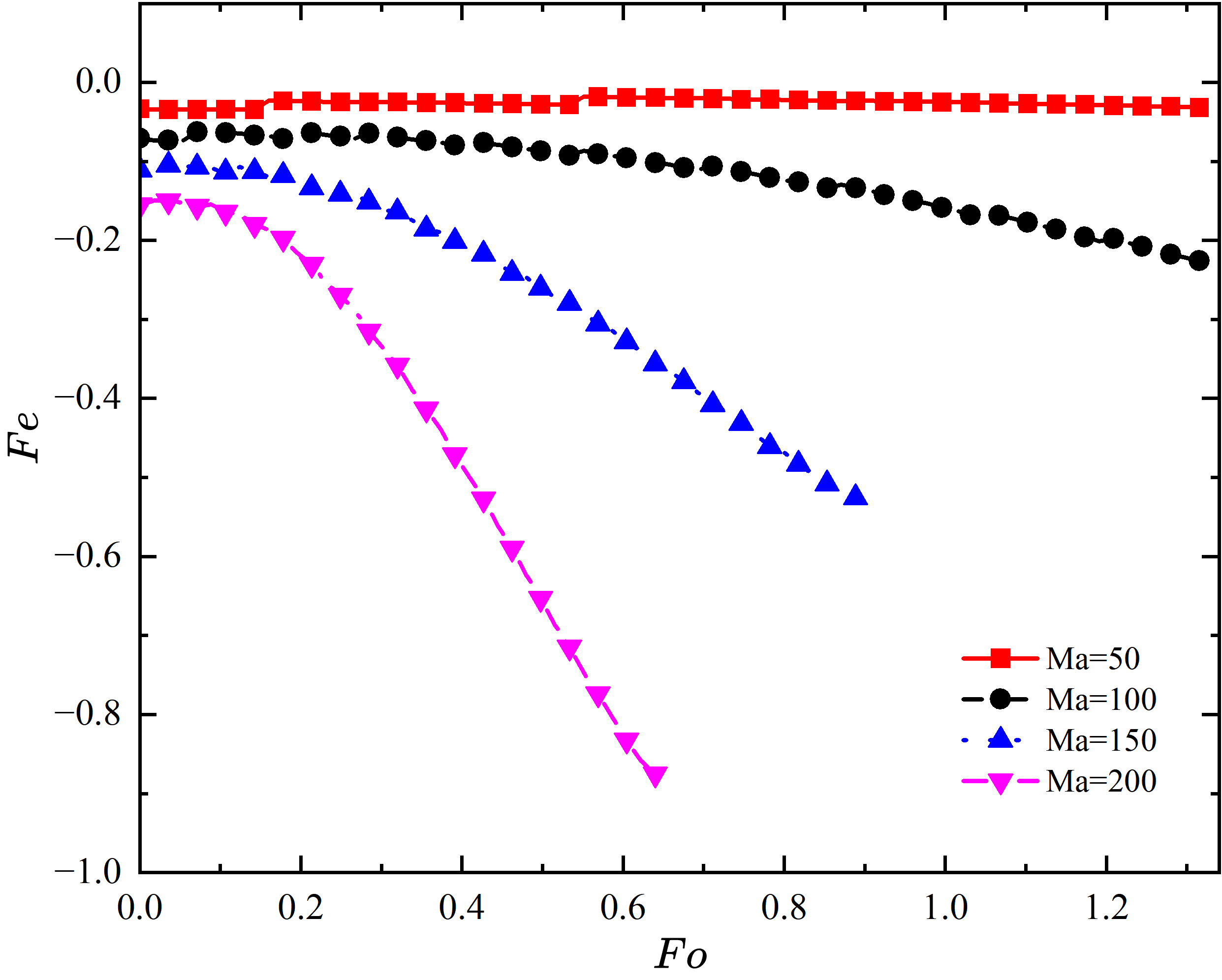}
	\caption{Relationship between the resultant force ${F_e}$ and ${F_o}$ for droplets at different Ma values, with ${F_e}$ normalized by surface tension.}
	\label{fig5-3}
\end{figure}
To provide a more intuitive reflection of the accuracy of the model in predicting droplet motion modes, we numerically study the evolution of the droplet's contact line length ${l_w}$ and centroid position ${x_p}$ with ${F_o}$, as depicted in Fig. 8. Under different inclinations of the inclined plane, the overall trend of the droplet's contact line length variation remains consistent, i.e., the contact line length increases with increasing Marangoni number. When the Marangoni numbers are $150$ and $200$, the rate of increase in contact line length is not very significant due to the rapid migration of the droplet reaching the boundary prematurely. However, when the Marangoni number is $100$, it can be observed that the change in contact line length gradually increases with the droplet's migration. Additionally, the relationship between the change in contact line length and ${F_e}$ remain consistent, as depicted in Fig. 7. As ${F_e}$ decreases negative, the variation in contact line length ${l_w}$ becomes more pronounced. When the Marangoni number is $50$, the trend of droplet change is not apparent. This aligns with the situation where the resultant force ${F_e}$ is close to zero, which represents the boundary for droplet length deformation and extension as shown in Fig. 5. Moreover, considering the relationship between the droplet's centroid position and length under different inclinations, the influence of gravity on this deformation and extension of the droplet is significant. When the droplet is positioned at the same location on the inclined plane, the variation in the droplet's contact line length due to changes in the inclination is not significant. The influence of gravity on the droplet primarily manifests in affecting its migration rate, which is why gravity is not considered as a part of the forces in the model discussed above.
\begin{figure}[H]
	\centering
	\subfigure[]{\label{fig5-4a}
		\includegraphics[width=0.4\textwidth]{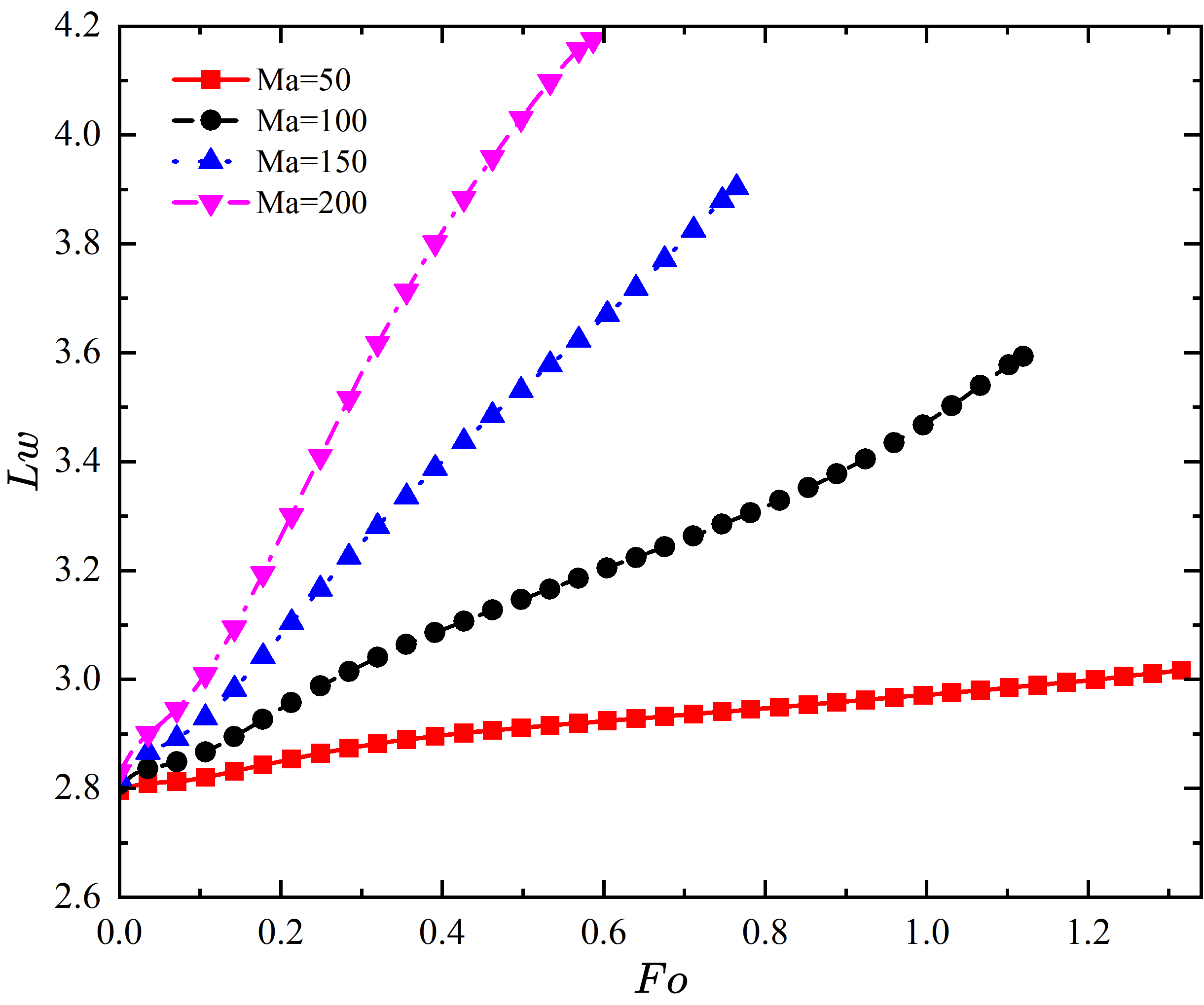}
	}
	\subfigure[]{\label{fig5-4b}
		\includegraphics[width=0.4\textwidth]{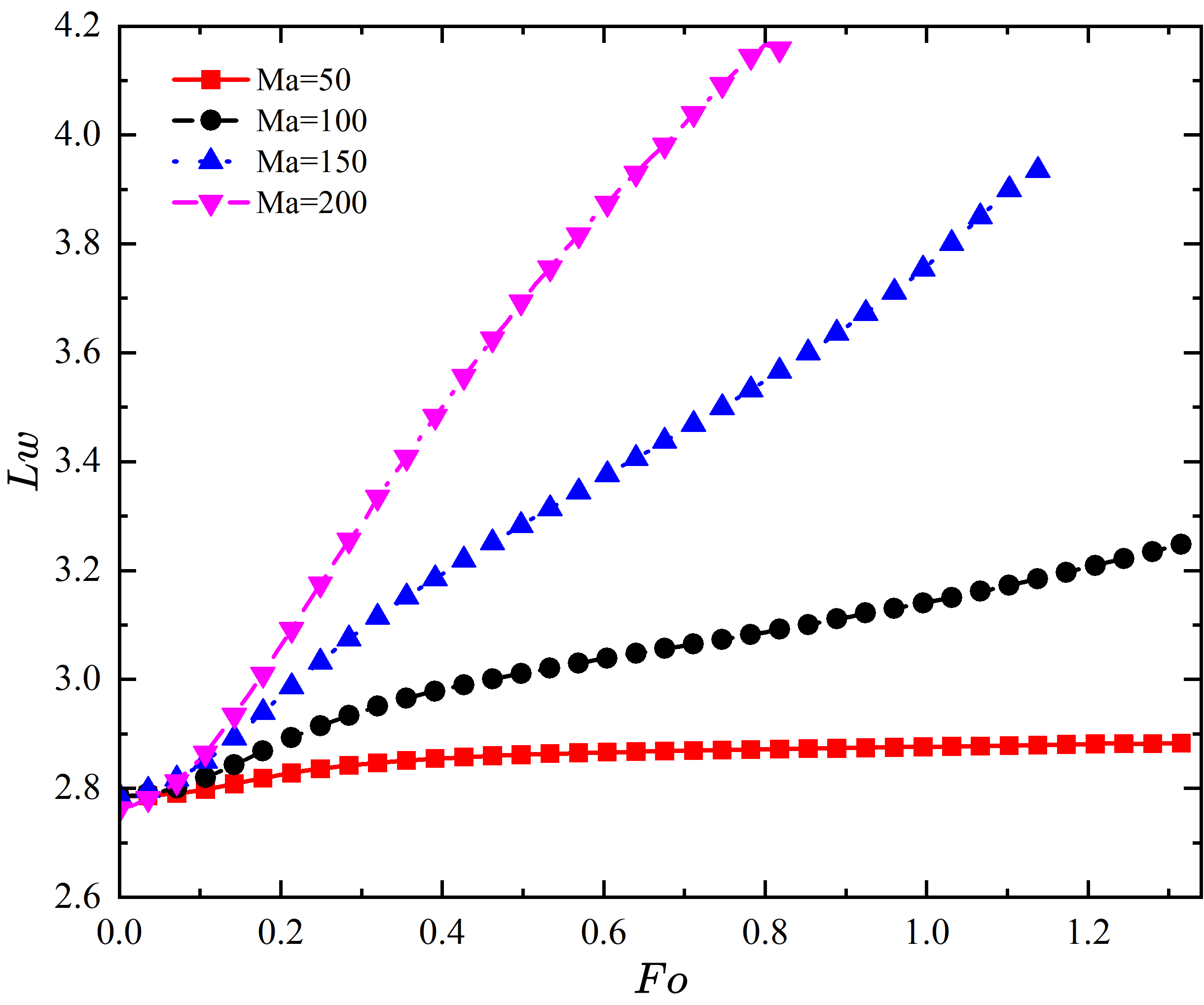}
	}
    \subfigure[]{\label{fig5-4d}
    	\includegraphics[width=0.4\textwidth]{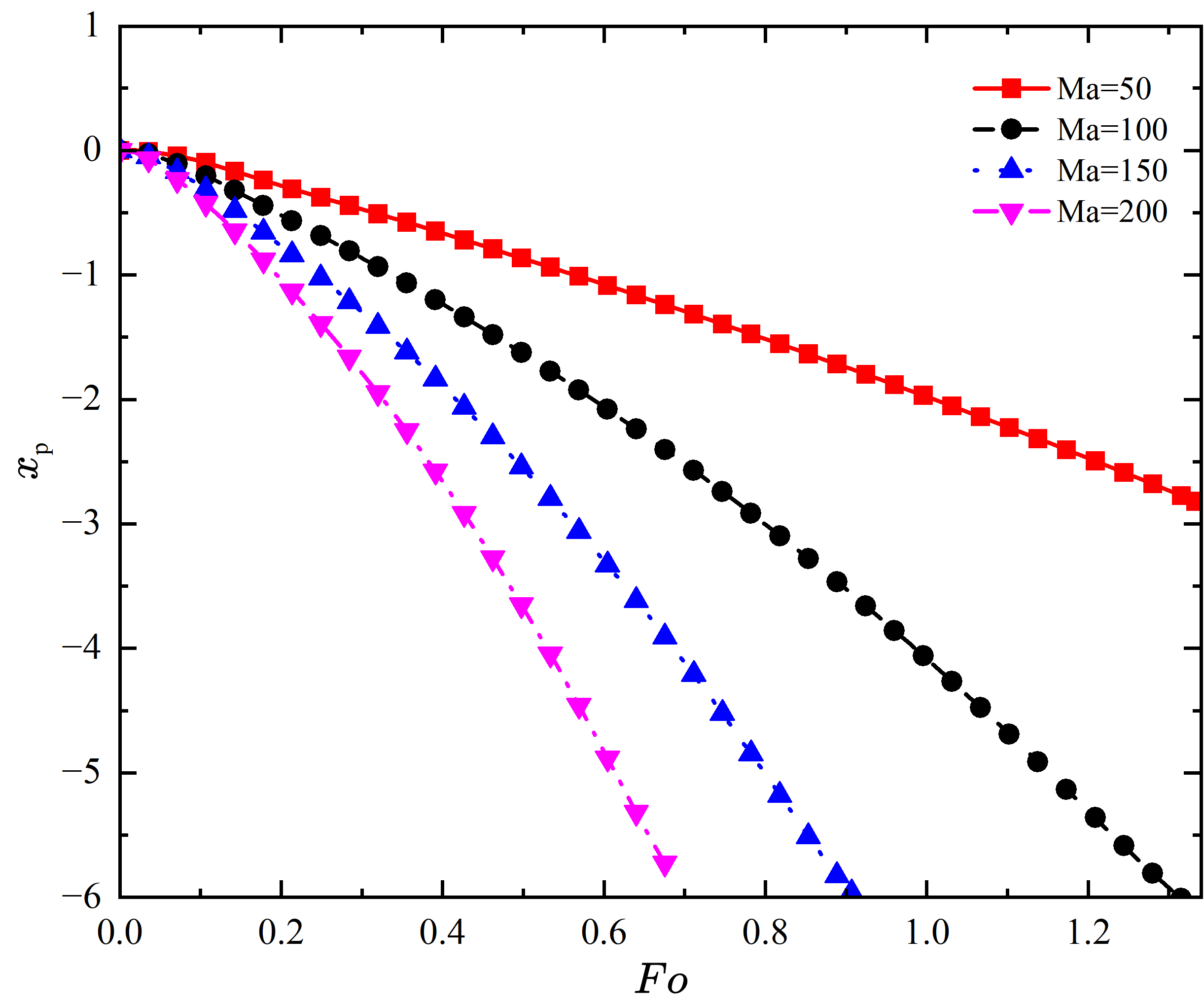}
    }
    \subfigure[]{\label{fig5-4f}
    	\includegraphics[width=0.4\textwidth]{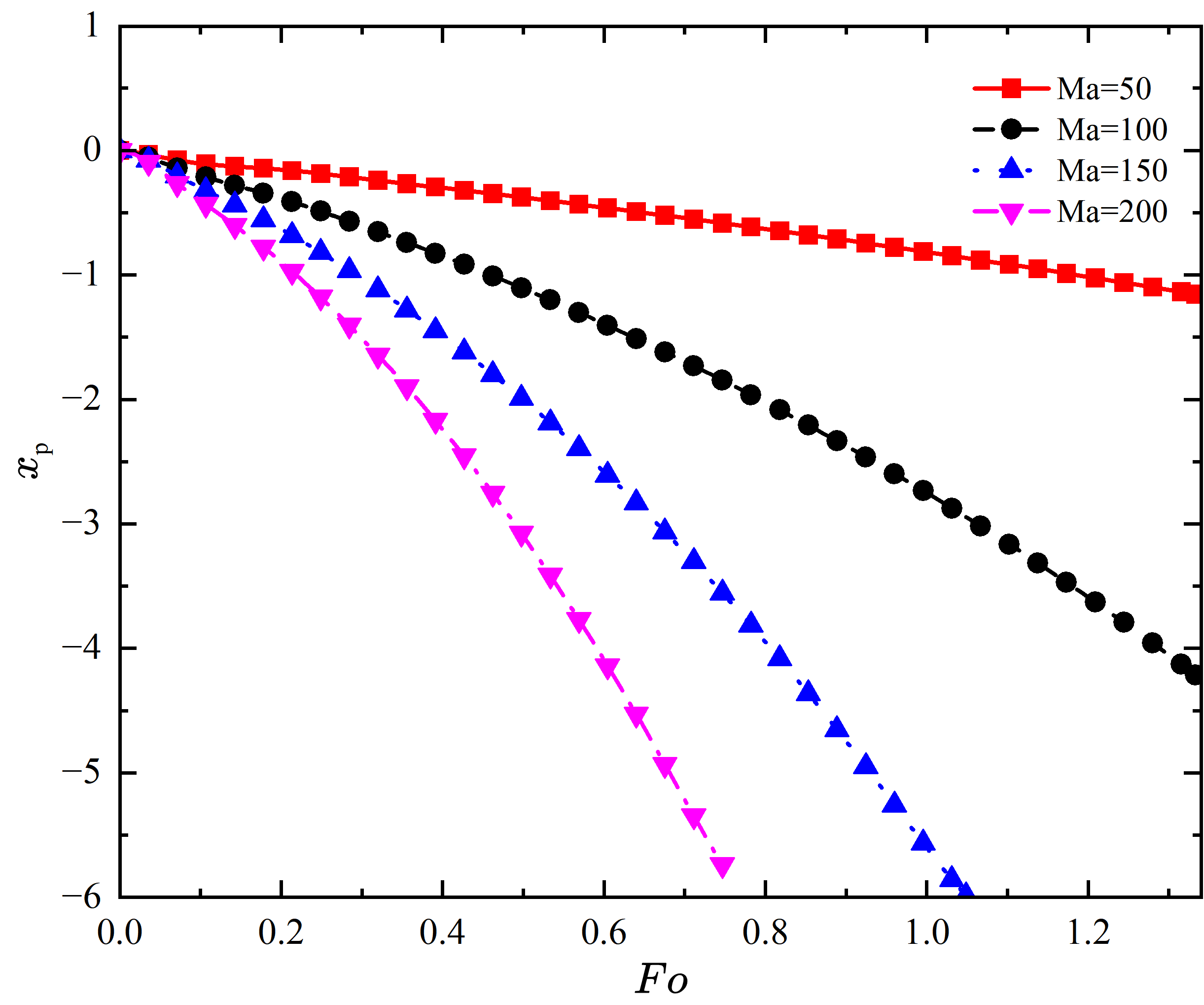}
    }
	\label{fig5-4}
	\caption{The relationship between contact line (${l_w}$) (normalized by the initial radius of the droplet), centroid position(${x_p}$)(normalized by the initial position) and ${F_o}$ at different titl angles: (a), (c) $\alpha  =  - 30^\circ $, (b), (d) $\alpha  =  30^\circ $.}
\end{figure}

\subsection{The impact of surface wettability on droplet migration}
In this subsection, the influence of surface wettability on droplet migration will be investigated. Due to the relatively large surface-to-volume ratio of droplet, the effect of surface wettability cann't be ignored, which is achieved by altering the contact angle $\theta $. We explore the effects of varying surface wettability by setting $\theta  = 60^\circ  \sim 120^\circ $, while keeping other parameters constant. For all contact angles, the droplet's motion on the inclined surface is primarily influenced by gravity and thermocapillary forces induced by temperature gradients on the surface. We initially conduct numerical simulations on the migration of self-rewetting droplets on flat surfaces. Similar to normal fluids \cite{Baroud_PRE2007}, the movement direction of self-rewetting droplets also changes with the contact angle. On hydrophilic surfaces, the droplets move towards the high surface energy regions and move towards the low surface energy regions while on hydrophobic surfaces. However, unlike normal droplet, the migration behavior of self-rewetting droplet is more diverse. Due to temperature differences at various locations, the migration speed of self-rewetting droplet changes, which makes their migration more controllable. On inclined surfaces, the droplet exhibits climbing against gravity due to the differing hydrophilicities. This can be explained by a simple force analysis considering the unbalanced surface tension and gravity acting on the droplet \cite{Gennes_RMP1985}
\begin{equation}
	{F_d} = {({\sigma _{SB}} - {\sigma _{SA}})_L} - {({\sigma _{SB}} - {\sigma _{SA}})_R} - \left( {{\rho _A} - {\rho _B}} \right)gS\sin \alpha  = {\sigma _L}\sin {\theta _L} - {\sigma _R}\sin {\theta _R} - \left( {{\rho _A} - {\rho _B}} \right)gS\sin \alpha, 
\end{equation}
where L and R represent the left and right contact points of the droplet with the solid surface, respectively. Initially, the ${T_m}$ is set to $20$, the surface tension on the left side of the droplet is always greater than on the right side. Since the overall morphology of the droplet is nearly symmetrical during migration, the left and right contact angles are approximately equal, ${\theta _L} \approx {\theta _R}$, for hydrophilic surfaces, ${\sigma _L}\sin {\theta _L} > {\sigma _R}\sin {\theta _R}$, causing the droplet to move towards the left end. From Fig. 9, it can be observed that vortices form around the droplet in all cases (contact angles of $60^\circ $, $90^\circ $, and $120^\circ $, which represent hydrophilic, neutral, and hydrophobic surfaces, respectively). But they differ significantly, at the contact angle of $60^\circ $, a counterclockwise vortex appears on the left side of the droplet, while an external clockwise vortex is present. This is evident from the temperature distribution around the droplet, which shows that both hydrophilic and hydrophobic surfaces exhibit a counterclockwise temperature gradient on the left half of the droplet. Marangoni stress causes fluid to flow from regions of lower interfacial tension. Due to the self-rewetting properties of the droplet, the interfacial tension in the high-temperature region is higher, resulting in flow directions opposite to those in normal fluids. On hydrophilic surfaces, the smaller curvature of the droplet leads to the formation of only one vortex. In contrast, on neutral or hydrophobic surfaces, the temperature gradient at the right end of the droplet causes a pair of counter-rotating vortices. Interestingly, previous lubrication theories suggested that fluid always moves toward the colder end under thermalcapillary action. However, due to the unique surface tension characteristics of self-rewetting fluids, which could lead the droplet move to the heated surface.
\begin{figure}[H]
	\centering
	\includegraphics[width=0.8\textwidth]{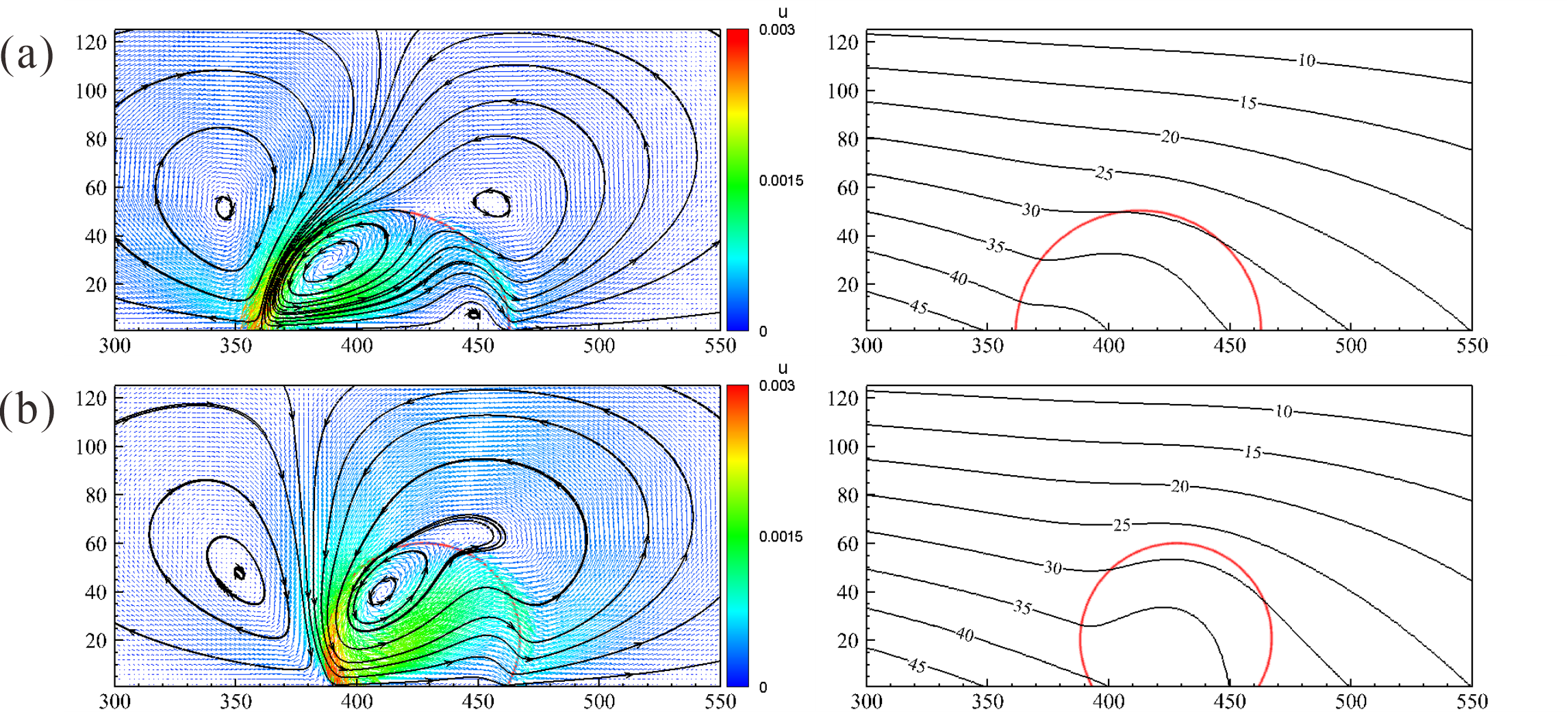}
	\caption{The flow field (the left plane) and the temperature field (the right plane) surrouding the moving droplet at different contact angles: (a) $\theta  = 90^\circ $, (b) $\theta  = 120^\circ $, in which the red lines are $\phi  = 0$, the colorful lines with arrows are the velocity vectors, and the black lines with arrows are the streamlines.}
	\label{fig6-1}
\end{figure}
Additionally, for hydrophilic surface, there is only one vortex inside the droplet, which increases in size as the inclination angle of the tilted substract increases and whose center gradually moves upward, Furthermore, for hydrophobic surface, there are two vortices inside the droplet, one on each side, where the left-side vortex also moves upward as the inclination angle increases, while the right-side vortexx, which gradually moves downward, eventually disappears. Considering the impact of the inclined plane, we also need to explore the influence of gravity on droplet motion at different contact angles. According to Eq.(31), the driving force of the droplet, ${F_d}$, can be derived: when ${F_d} > 0$, the droplet moves toward the hot end, while as ${F_d} < 0$, it moves toward the cold end. Therefore, for dofferemt contact angle, there exists a critical tilt angle of the inclined surface. If the tilt angle exceeds this value, the droplet moves toward the cold end, otherwise, it moves toward the hot end. Before calculating this critical angle, we must determine the contact line length of the droplet at various contact angle, which allows us to derive the value of the driving force ${F_d}$ for the droplet at each contact angle. Since the volume of the droplet remains consistent
\begin{equation}
	V = \frac{1}{2}\pi {R^2},
\end{equation}
in such a case, for a droplet with a contact angle of $\theta $, the base radius $r$ can be determined
\begin{equation}
	r = \sqrt {\frac{{0.5\pi }}{{\theta  - \sin \theta \cos \theta }}} R,
\end{equation}
as a result, the half-length $l$ of the base can be obtained
\begin{equation}
	l = r\sin \theta.
\end{equation}
In addition, for a droplet with a contact angle of $\theta $ moving on an inclined plane with an angle of $\alpha $, the driving force ${F_d}$ is given by
\begin{equation}
	{F_d} = \left( { - 2{\sigma _T}Gl + 2{\sigma _{TT}}Gl{T_h}} \right)\cos \theta  - ({\rho _A} - {\rho _B})gS\sin \alpha , 
\end{equation}
when ${F_d}$ equals zero, the droplet is in a critical state of motion. Therefore, the explicit expression of the tilt angle $\alpha $ and the contact angle $\theta $ can be derived
\begin{equation}
	\alpha  = \arcsin \left[ {{{\left( {2G\sqrt {\frac{{0.5\pi }}{{\theta  - \sin \theta \cos \theta }}} R\sin \theta \cos \theta \left( { - {\sigma _T} + {\sigma _{TT}}{T_h}} \right)} \right)} \mathord{\left/
				{\vphantom {{\left( {2G\sqrt {\frac{{0.5\pi }}{{\theta  - \sin \theta \cos \theta }}} R\sin \theta \cos \theta \left( { - {\sigma _T} + {\sigma _{TT}}{T_h}} \right)} \right)} {({\rho _A} - {\rho _B})gS}}} \right.
				\kern-\nulldelimiterspace} {({\rho _A} - {\rho _B})gS}}} \right].
\end{equation}
\begin{figure}[H]
	\centering
	\includegraphics[width=0.6\textwidth]{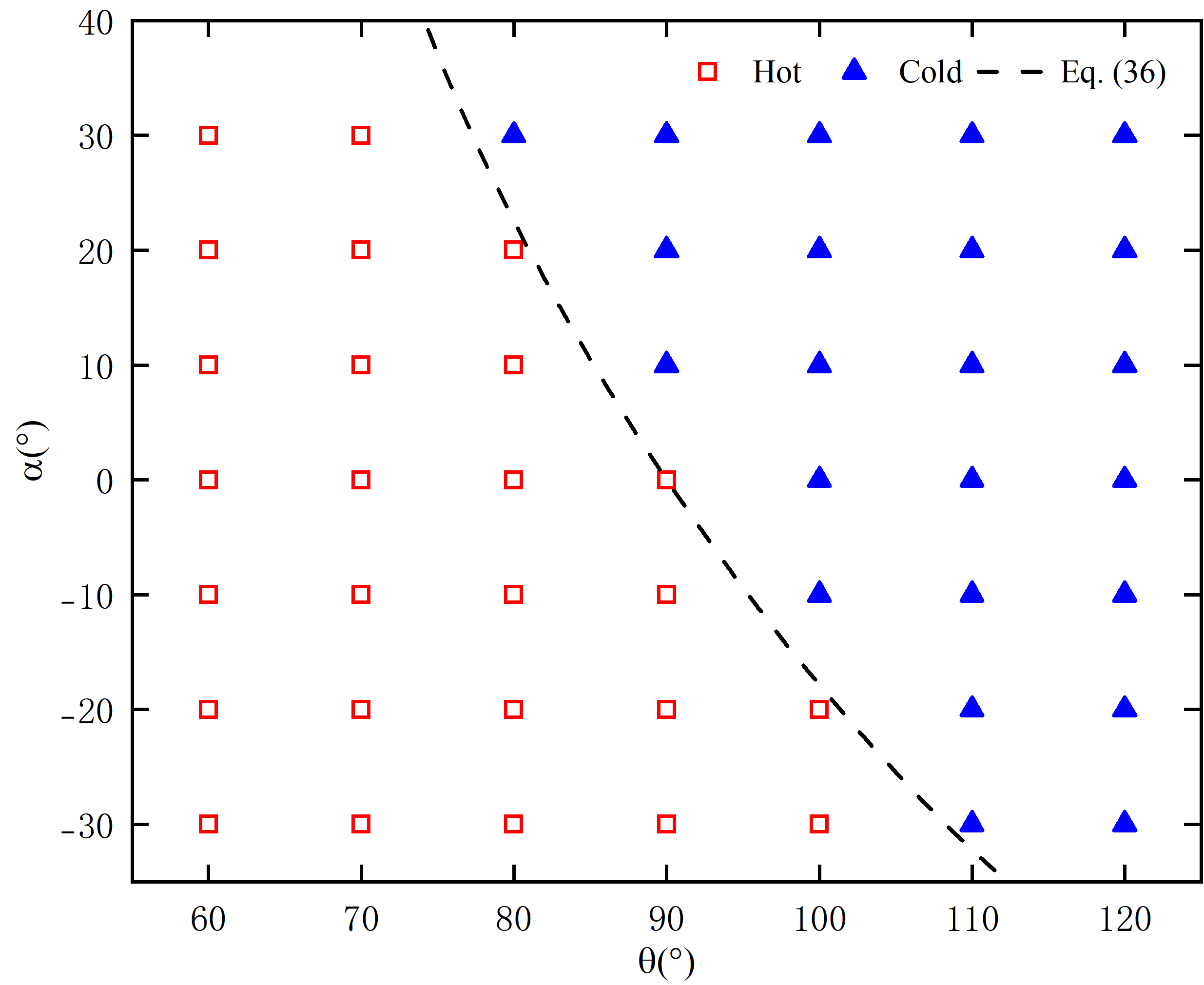}
	\caption{The horizontal axis represents the contact angle of the droplet, and the vertical axis represents the tilt angle of the inclined plate. Red squares indicate droplets moving towards the hot end, blue triangles indicate droplets moving towards the cold end, and the black dashed line represents the theoretical critical value for movement towards the hot end and cold end.}
	\label{fig6-2}
\end{figure}
To valid this theory, we simulate the thermocapillary migration of the self-rewetting droplet under different contact angles and tilt angles. The results are shown in Fig. 10, where the blue squares and red triangles represent the droplet moves to the cold end and the hot end, respectively. The dashed line stands for the critical state of droplet motion derived from Eq. (36), where the droplet remains stationary. It can be seen that more hydrophilic or hydrophobic surfaces increase the driving force on the droplet, which allows it to climb against gravity at larger tilt angles.

To explore the impact of droplet elongation on deformation under hydrophilic and hydrophobic conditions, we also examined the migration of droplets with contact angles ranging from $30^\circ $ to $150^\circ $ in different directions on inclined planes. As shown in Fig. 11, the stronger the hydrophilicity or hydrophobicity of the droplet, the faster the migration speed. However, the effect is not significant for hydrophobic surfaces. This is because the droplet volume remains constant as the contact angle alters, and the contact line length shortens, reducing the surface tension difference on both sides and gradually weakening the driving force of the droplet. We also investigated the initial force on the droplet, finding that the normalized net force ${F_e}$ is approximately -0.1 for a contact angle of $30^\circ $, while it approaches zero for contact angles of $60^\circ $, $90^\circ $, and $120^\circ $. From the figure, it is also clear that the length of the droplet significantly extends at a contact angle of $30^\circ $, while it remains nearly balanced in other states. Interestingly, at a contact angle of $150^\circ$, the droplet shortens because the large wetting angle generates an excessive horizontal inward component of surface tension, which causes the droplet to contract. This finding fits well with theoretical predictions. Additionally, this deformation is not directly related to gravity. Instead, gravity affects the droplet's migration speed and direction, indirectly causing different droplet modes.
\begin{figure}[H]
	\centering
	\subfigure[]{\label{fig6-3a}
		\includegraphics[width=0.4\textwidth]{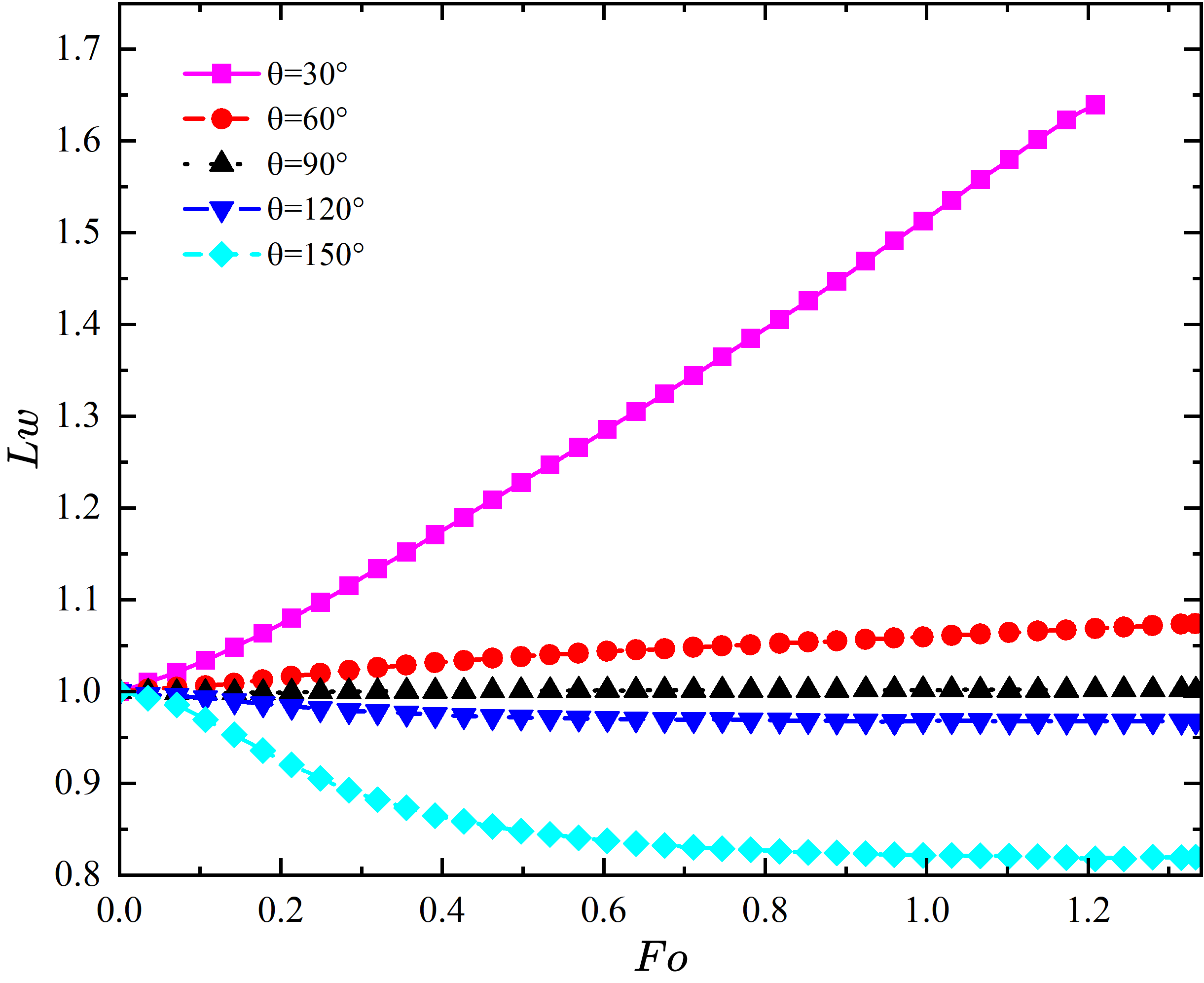}
	}
	\subfigure[]{\label{fig6-3b}
		\includegraphics[width=0.4\textwidth]{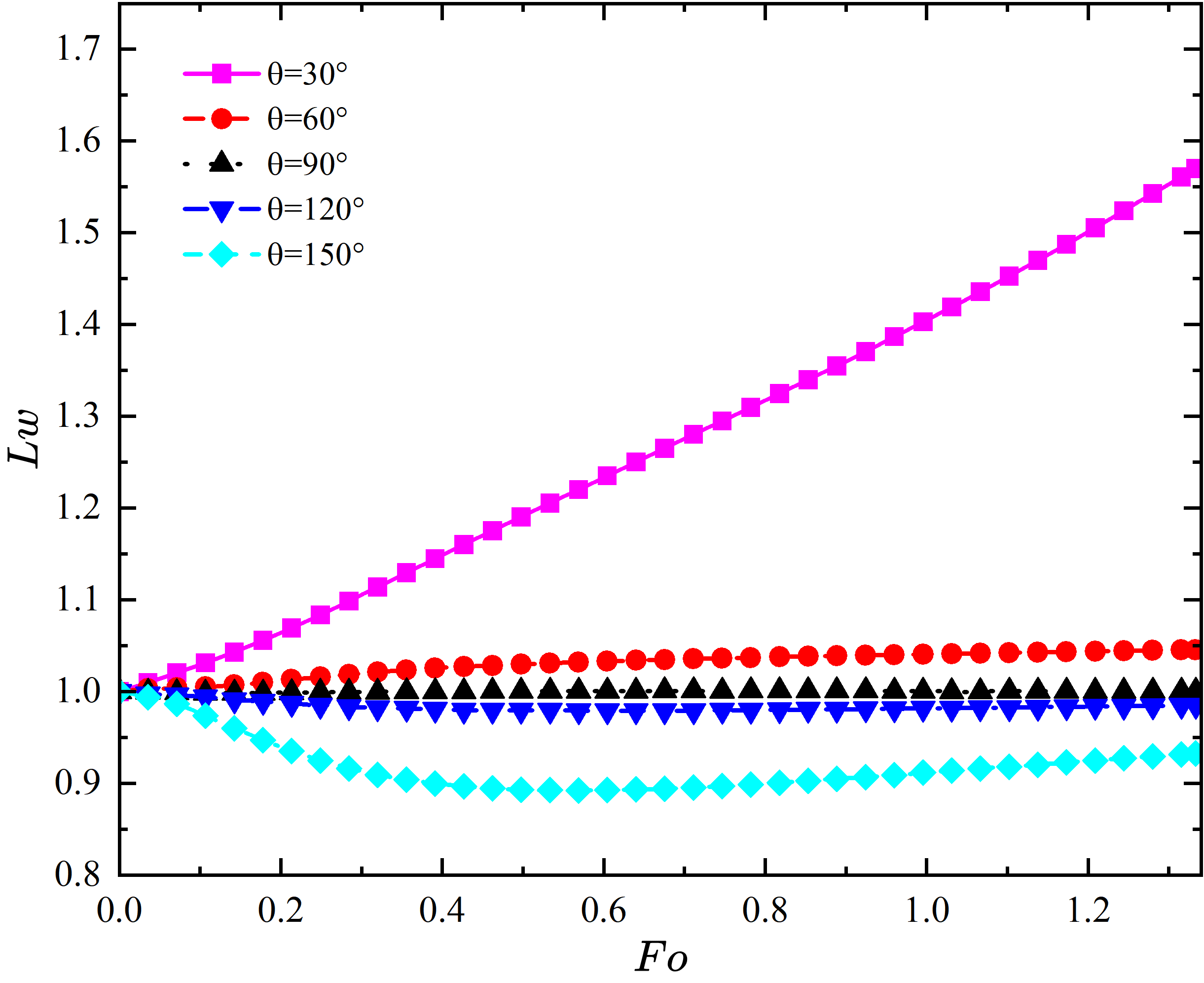}
	}
	\subfigure[]{\label{fig6-3d}
		\includegraphics[width=0.4\textwidth]{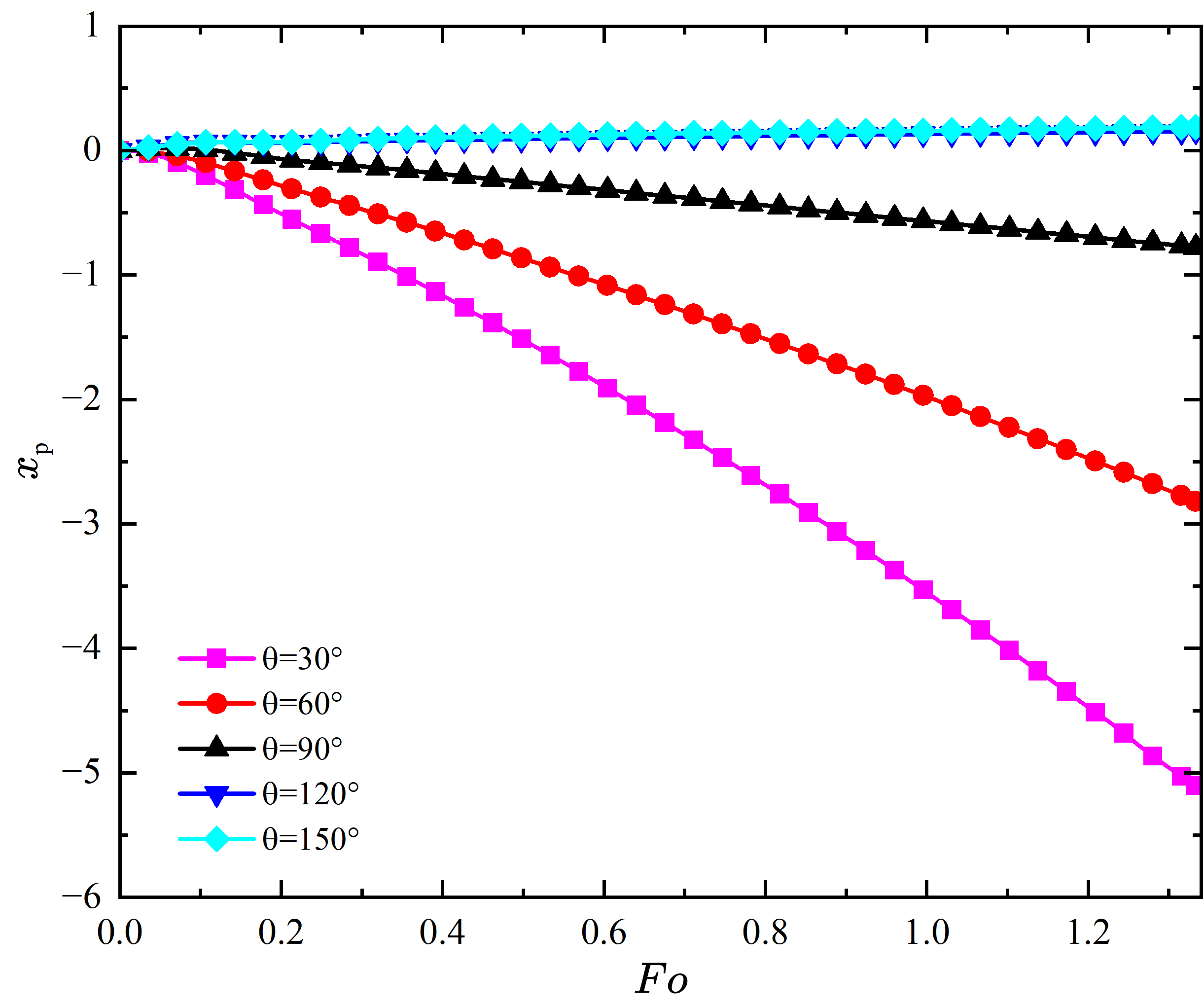}
	}
	\subfigure[]{\label{fig6-3f}
		\includegraphics[width=0.4\textwidth]{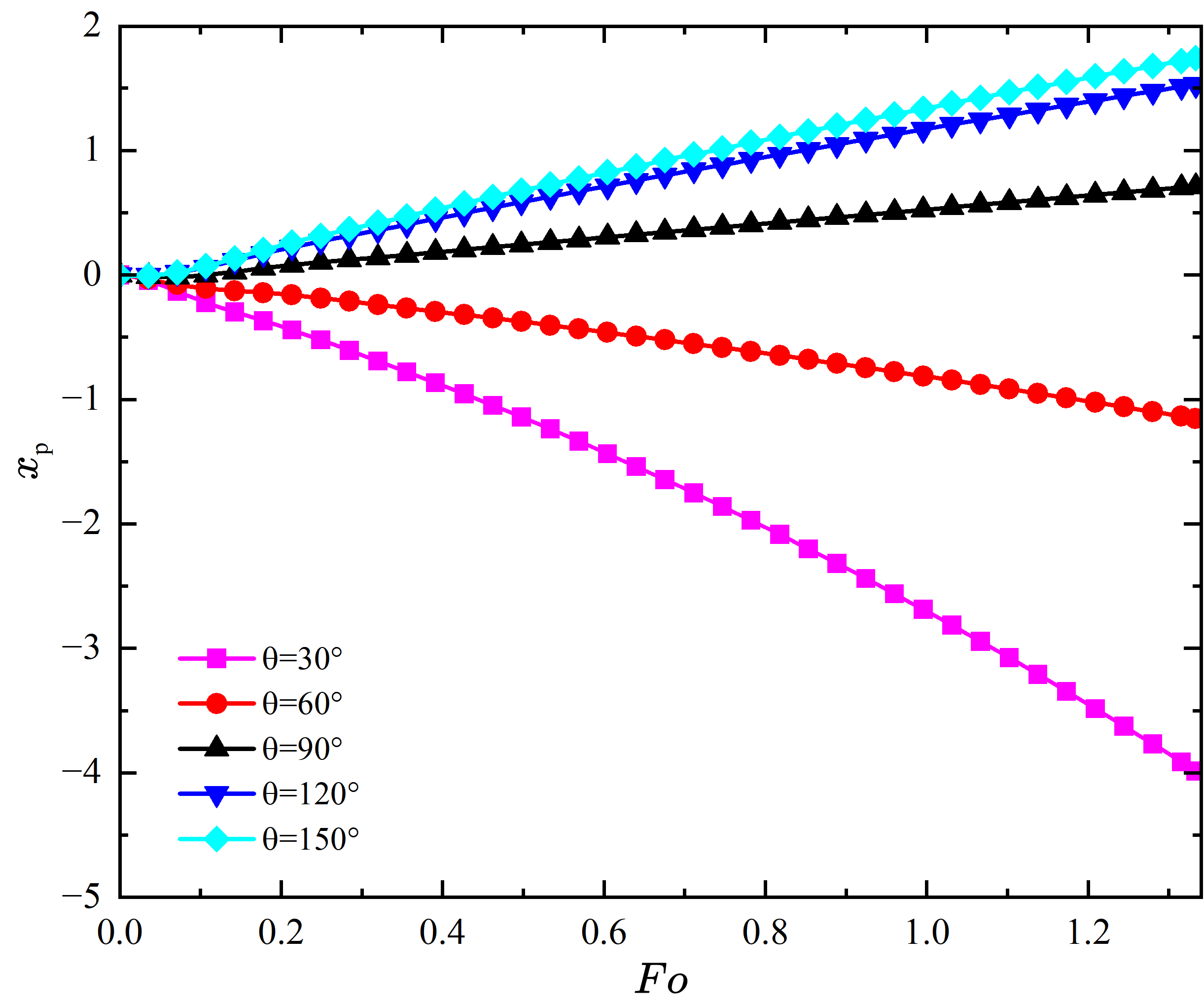}
	}
	\label{fig6-3}
	\caption{The relationship between contact line (${l_w}$), centroid position (${x_p}$) and ${F_o}$ at different titl angles: (a), (c) $\alpha  =  - 30^\circ $, (b), (d) $\alpha  =  30^\circ $.}
\end{figure}

\subsection{The impact of viscosity ratio on droplet migration}
The effect of viscosity ratio on droplet's thermocapillary migration will be investigated in this subsection. This is achieved by keeping the droplet viscosity constant while varying the viscosity of the external fluid to obtain different viscosity ratios. 
Fig. 12 shows the flow field and temperature field distribution around a droplet on a $30^\circ $ inclined plate at ${F_o} = 1.33$ for different viscosity ratios. Unlike normal fluids, the vortices in self-rewetting fluids exhibit significant changes due to viscosity differences. When the viscosity ratio is low (${{{\nu _B}} \mathord{\left/
		{\vphantom {{{\nu _B}} {{\nu _A}}}} \right.
		\kern-\nulldelimiterspace} {{\nu _A}}} \le 1$), 
only one vortex appears inside and outside the droplet. However, when the viscosity ratio is higher, an additional vortex appears on the right side of the droplet. This right-side vortex is influenced not only by temperature but also by gravity. Previous studies on normal fluids have found that changes in viscosity can also alter the droplet's movement direction \cite{Fath_IJMF2015}. However, in this simulation, an increase in viscosity only slows down the droplet's migration velocity, while the migration direction remains consistent across different tilt angles. For negative inclination angles where the droplet moves toward the lower part of the inclined plate, only one vortex appears regardless of the viscosity ratio. As the inclination angle increases, a vortex eventually forms on the right side of the droplet. The higher the viscosity ratio, the smaller the inclination angle required for vortex formation on the right side of the droplet. Fig. 13 depicts the phase diagram of viscosity ratio versus inclination angle. Additionally, the size of the clockwise vortex increases with the viscosity ratio increased.
\begin{figure}[H]
	\centering
	\includegraphics[width=0.8\textwidth]{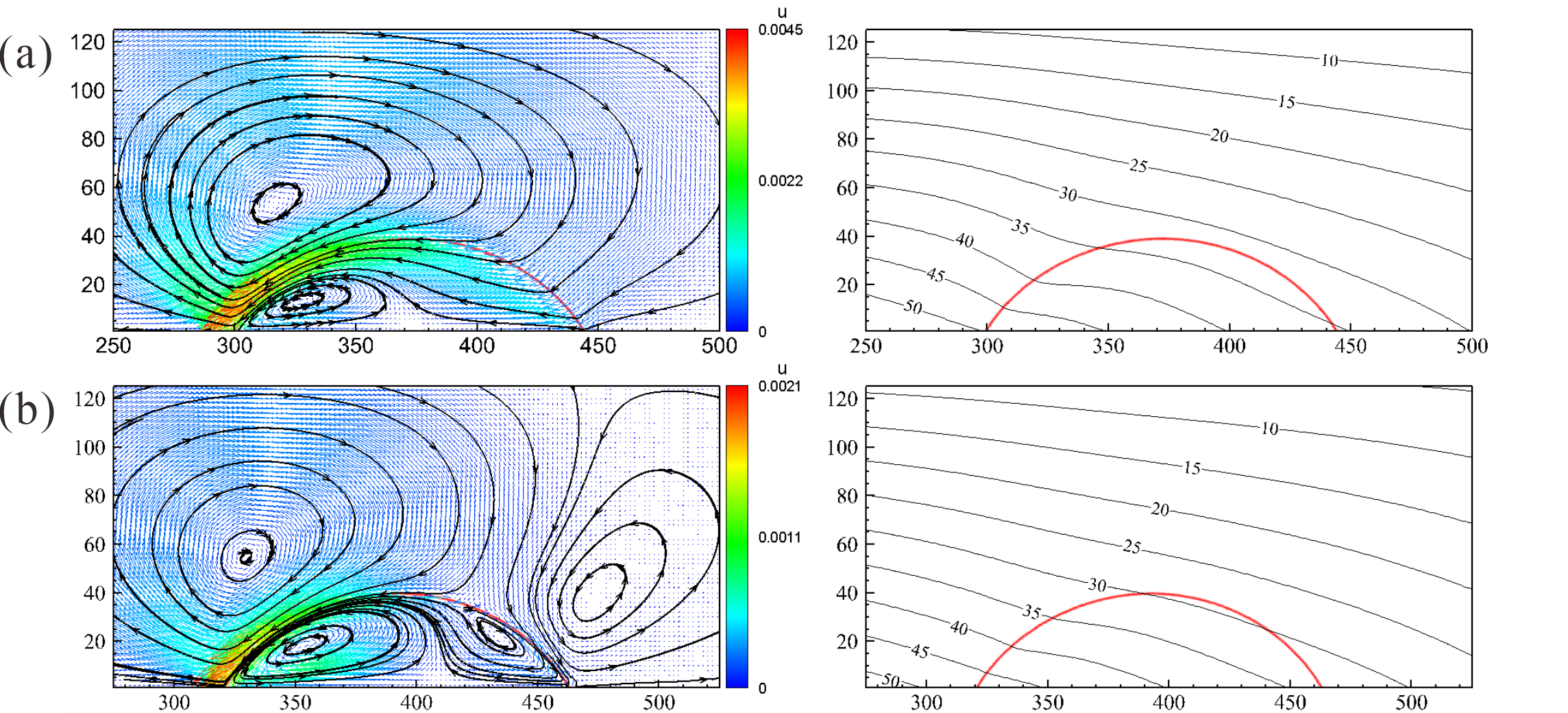}
	\caption{The flow field (the left plane) and the temperature field (the right plane) surrouding the moving droplet at the contact angle $\theta  = 60^\circ $ (a)${{{\nu _B}} \mathord{\left/
				{\vphantom {{{\nu _B}} {{\nu _A}}}} \right.
				\kern-\nulldelimiterspace} {{\nu _A}}} = 0.35$, (b)${{{\nu _B}} \mathord{\left/
				{\vphantom {{{\nu _B}} {{\nu _A}}}} \right.
				\kern-\nulldelimiterspace} {{\nu _A}}} = 3.5$. The red lines are $\phi  = 0$, the colorful lines with arrows are the velocity vectors, and the black lines with arrows are the streamlines.}
	\label{fig7-1}
\end{figure}
\begin{figure}[H]
	\centering
	\includegraphics[width=0.6\textwidth]{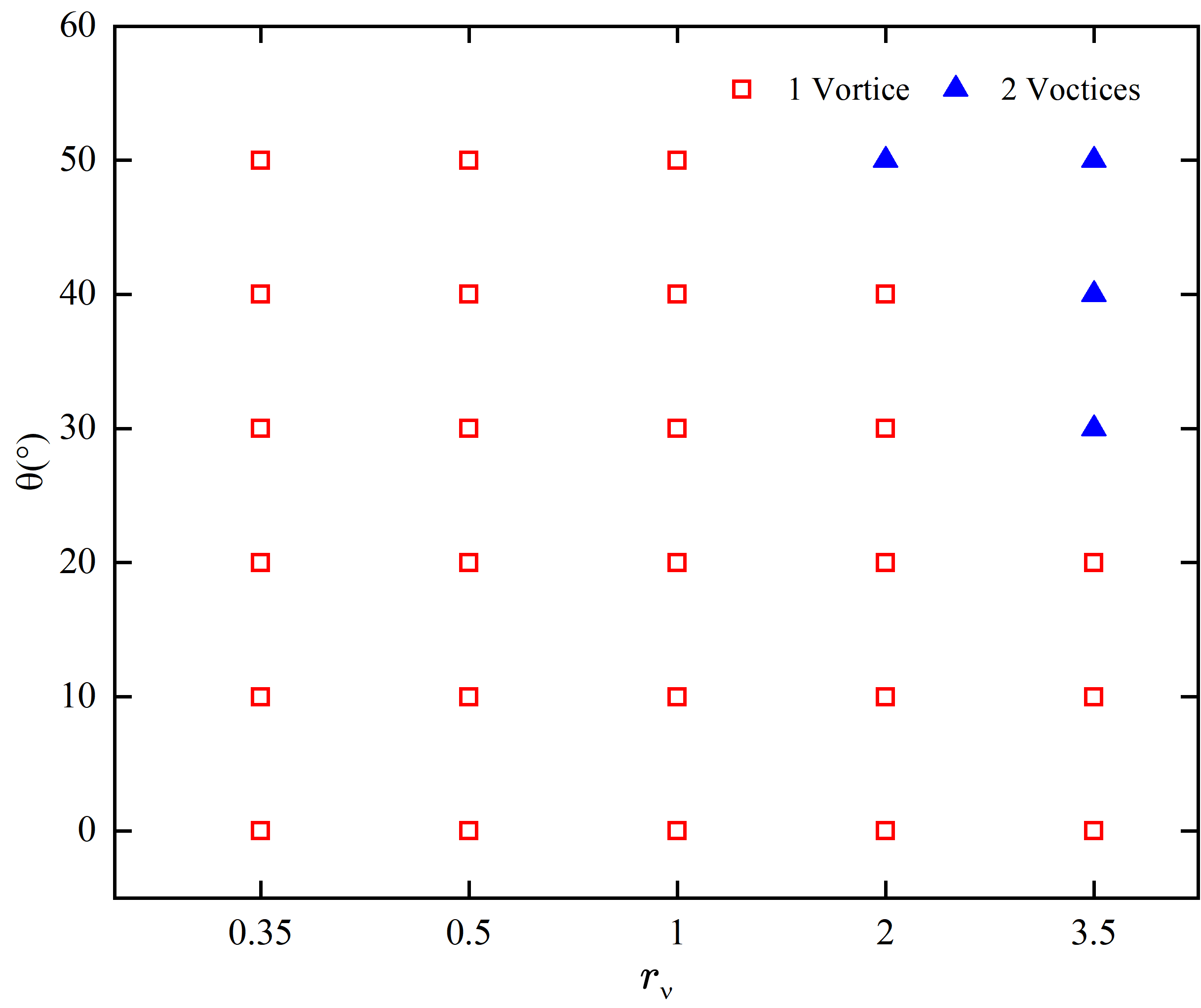}
	\caption{Phase diagram: The horizontal axis represents the viscosity ratio between the droplet and the fluid, and the vertical axis represents the tilt angle of the inclined plate. Red squares indicate a single vortex inside the droplet, and blue triangles indicate two vortices inside the droplet.}
	\label{fig7-2}
\end{figure}
Similarly, we track the net force on droplets under different viscosity ratios. Since the contact angle is maintained at $60^\circ $, the net force  ${F_e}$ on the droplet is always negative. As the viscosity ratio increases, the absolute value of  ${F_e}$ decreases. However, compared to changes in the Marangoni number, the viscosity ratio has a less noticeable effect on droplet elongation due to the small value of ${F_\mu }$ in the net force. The viscosity ratio mainly affects droplet migration speed. To better understand the changes in droplet migration behavior, we numerically study the relationship between the droplet centroid and contact line length with ${F_o}$ under different inclination angles (Fig. 14). The droplet migration speed decreases with increasing viscosity ratio because of the increased viscous shear resistance on the droplet. Under the set viscosity ratios in this study, no different migration directions of droplets due to varying viscosity ratios were observed. their movement directions remain consistent at the same inclination angle. Although the viscosity ratio influences droplet migration speed, the droplets still exhibit accelerated motion, as indicated by the slope of the centroid graph. The change in contact line length becomes less significant with increasing viscosity ratio. During the uphill movement of the droplet, its driving force counters gravity and viscous resistance, leading to slower migration and less noticeable elongation of the contact line.
\begin{figure}[H]
	\centering
	\subfigure[]{\label{fig7-3a}
		\includegraphics[width=0.4\textwidth]{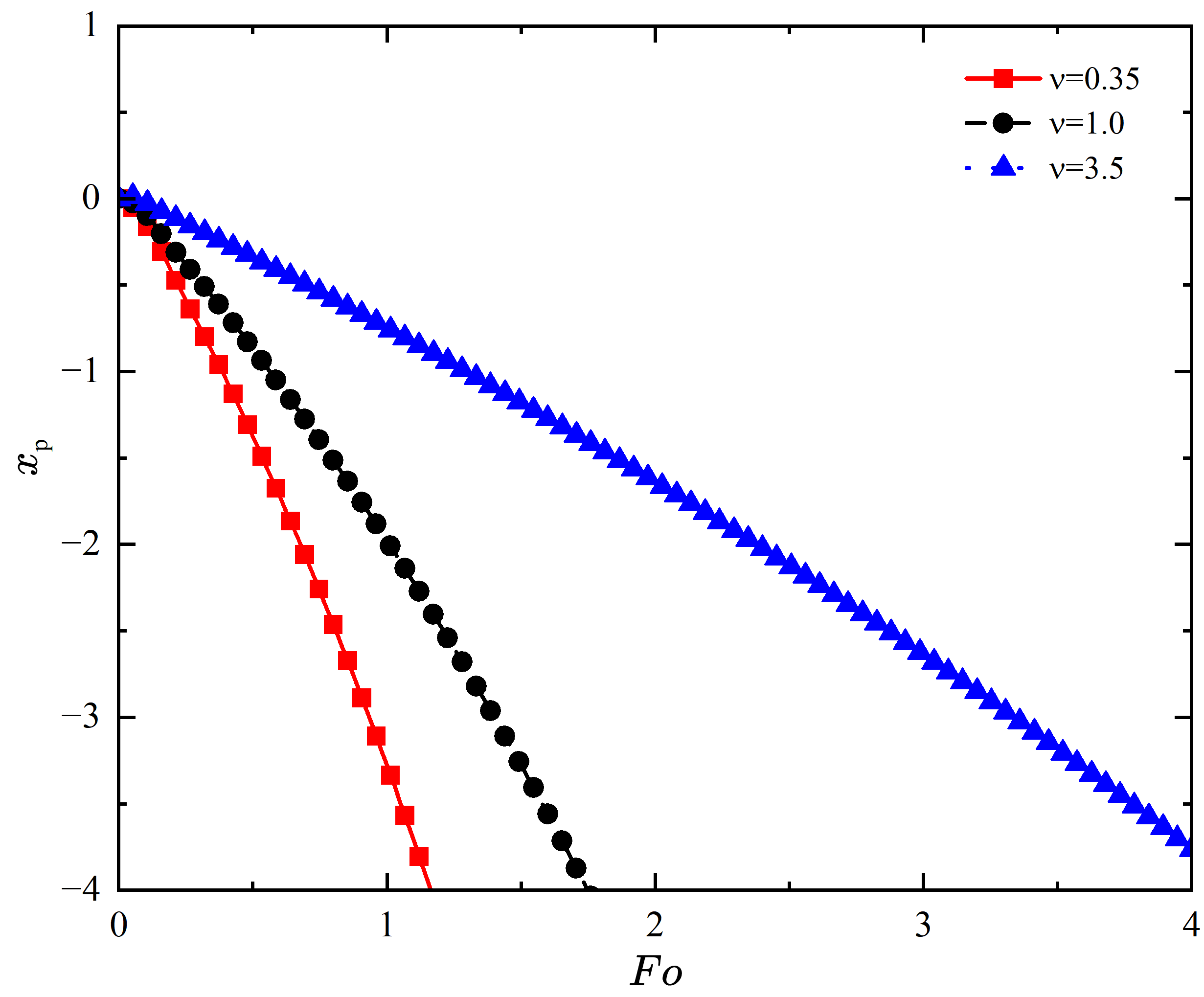}
	}
	\subfigure[]{\label{fig7-3b}
		\includegraphics[width=0.4\textwidth]{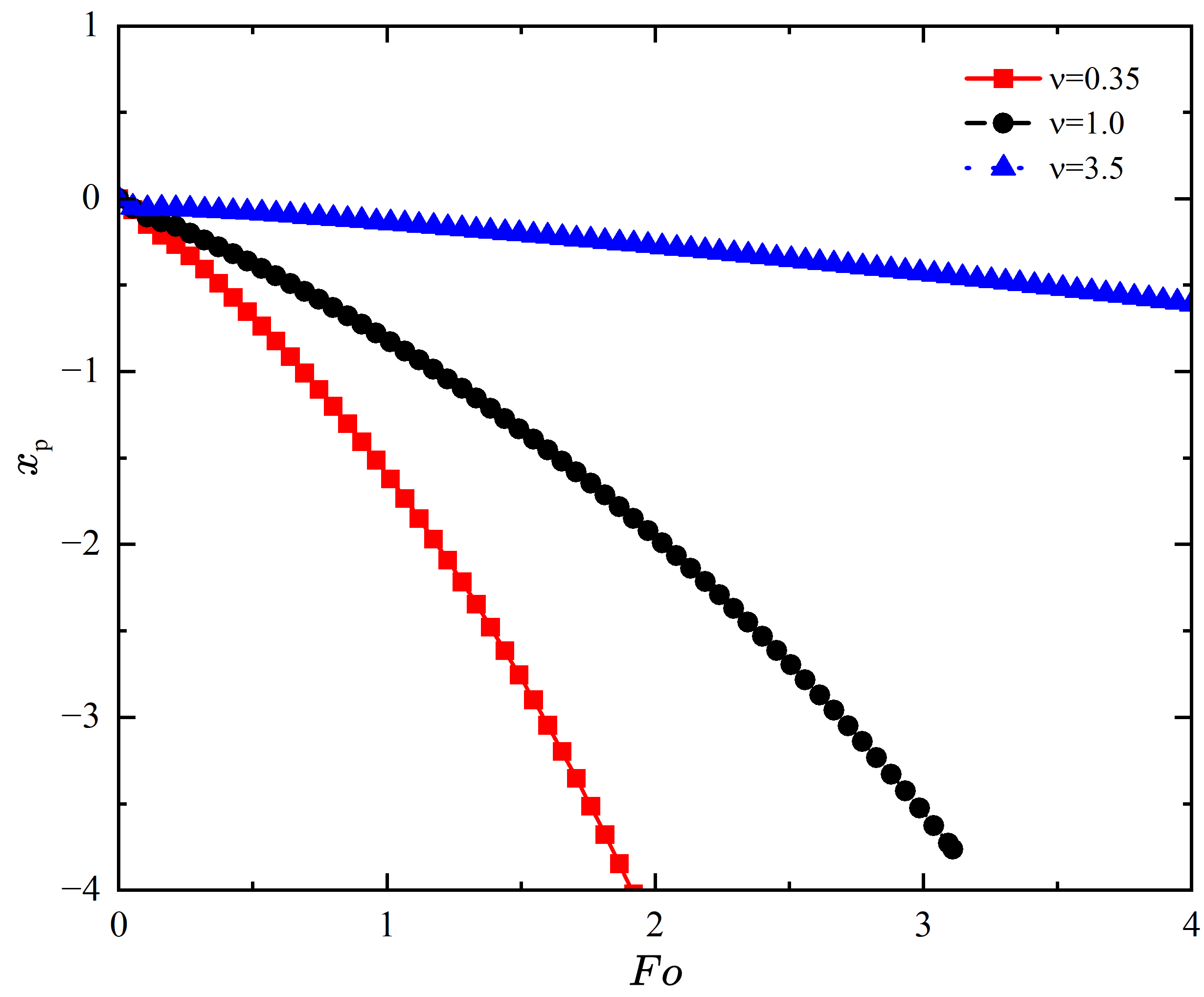}
	}
	\subfigure[]{\label{fig7-3d}
		\includegraphics[width=0.4\textwidth]{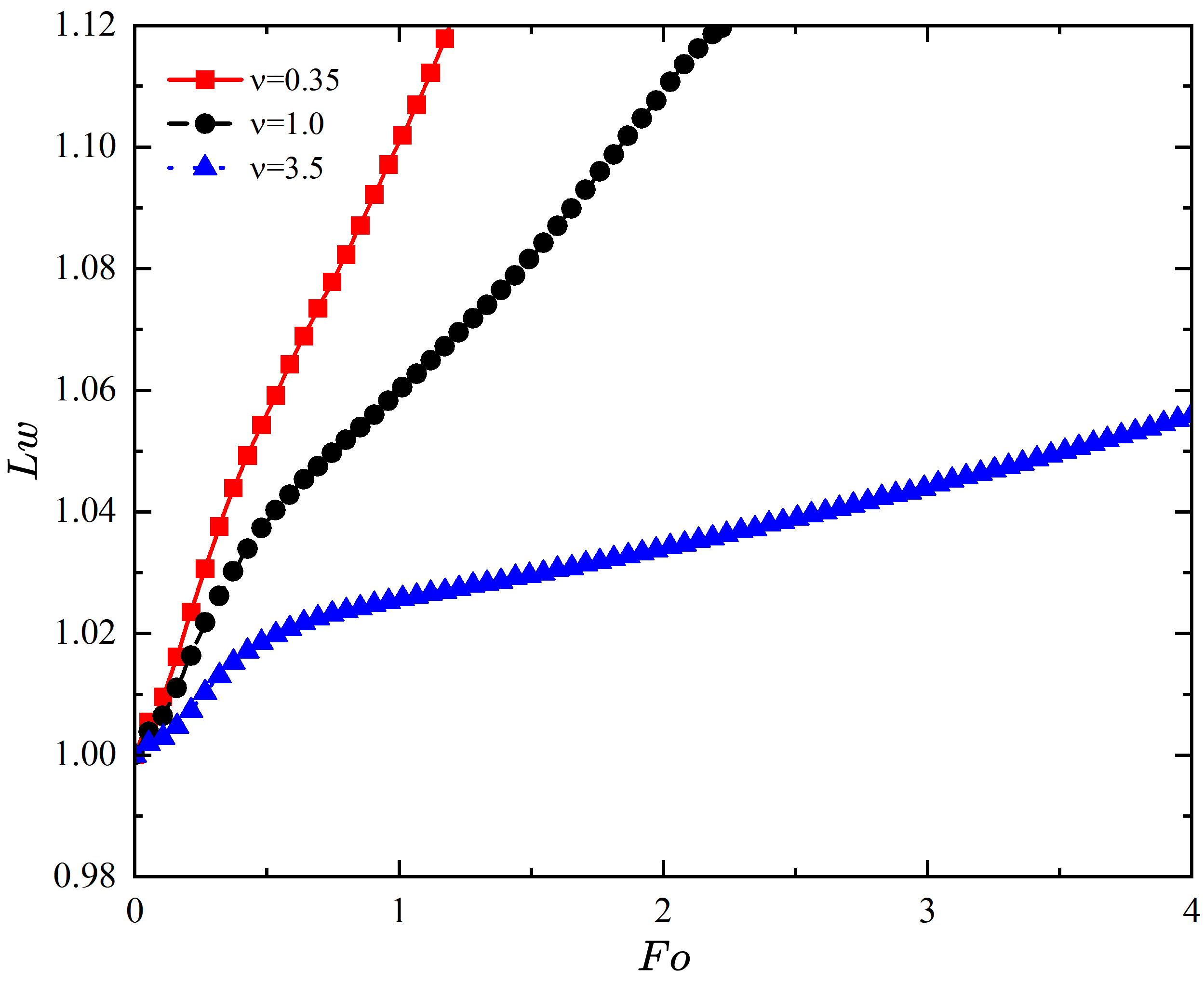}
	}
	\subfigure[]{\label{fig7-3f}
		\includegraphics[width=0.4\textwidth]{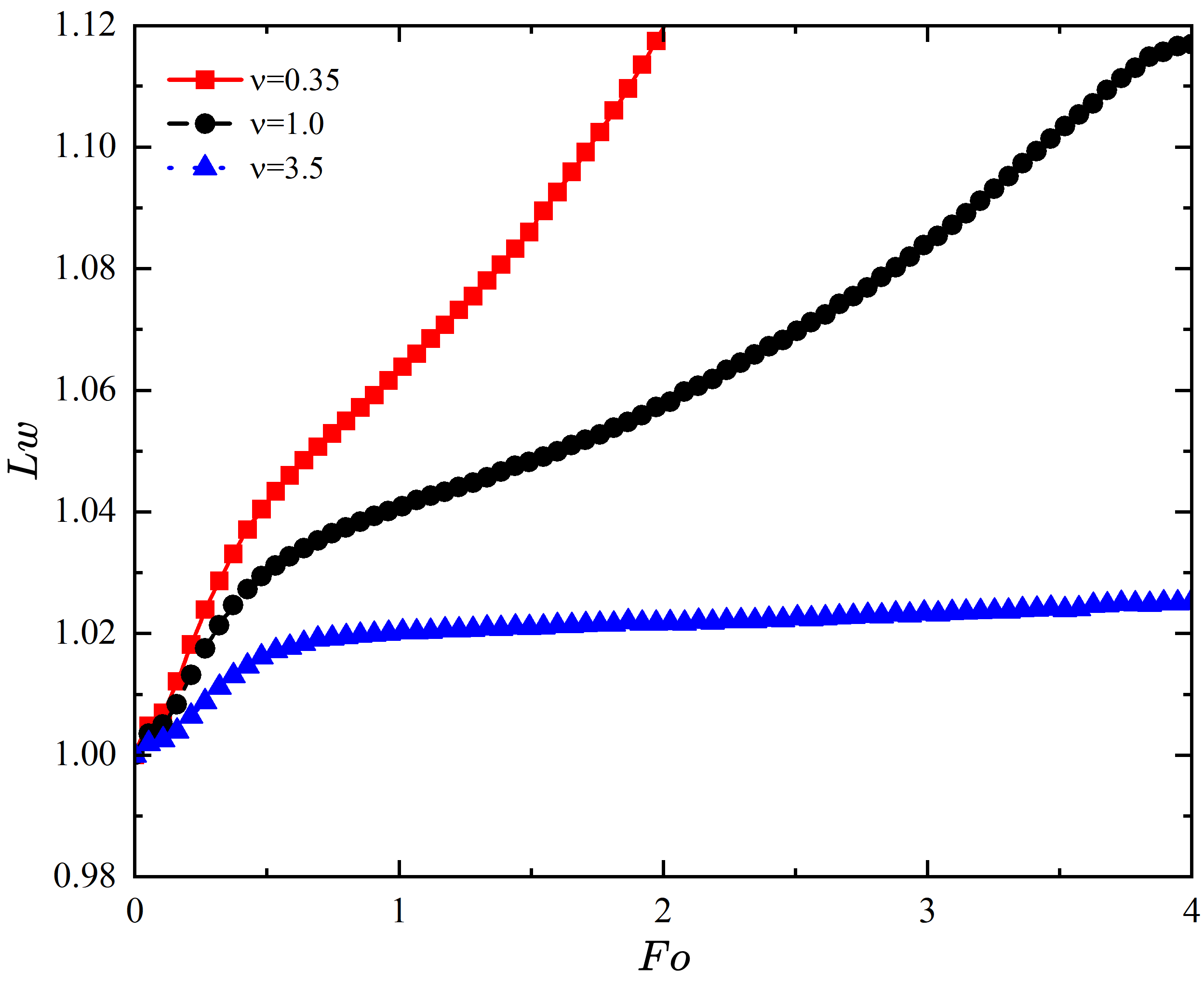}
	}
	\label{fig7-3}
	\caption{The relationship between contact line (${l_w}$), centroid position (${x_p}$) and ${F_o}$ at different titl angles: (a), (c) $\alpha  =  - 30^\circ $, (b), (d) $\alpha  =  30^\circ $.}
\end{figure}

\section{Conclusions}
Thermocapillary migration holds significant importance in scientific production, and in recent years, numerous experimental and numerical studies have focused on this phenomenon. However, previous research has predominantly considered the movement of droplets on temperature gradient surfaces for either normal fluids \cite{Fath_IJMF2015,Sui_POF2014}or self-rewetting fluids \cite{Xu_JFM2021} on flat surfaces. In practical industrial production, maintaining strictly horizontal conditions is challenging, especially in space and terrestrial thermal management devices, where microgravity and inclination are critical parameters that must be considered. Therefore, this study simulates thermocapillary flow based on the phase-field LB method, which can handle comparisons of thermal physical parameters.

Firstly, the thermocapillary migration of normal fluids is compared with previous methods to validate this numerical approach for handling solid-liquid interfaces. Subsequently, this model is used to simulate the thermocapillary migration of microfluidic droplets on a uniformly temperature-gradient inclined solid surface. The effects of different Marangoni numbers, surface wettability, and viscosity on droplet migration are explored. It is found that as the Marangoni number increases, the droplet gradually begins to deform and elongate, consistent with previous work and theoretical research. In addition, the migration direction of the droplet is influenced by different surface wettabilities. The migration direction changes under different wetting angles and tilt angles of the inclined plate, with variations in the internal vortices of the droplet observed. On hydrophilic surfaces, there is only one vortex inside the droplet, which increases in size as the inclination angle increases. On hydrophobic surfaces, however, there are vortices on both the left and right sides of the droplet. As the inclination angle increases, the left vortex rises while the right vortex moves downward. With further increases in the inclination angle, the right vortex eventually disappears. Moreover, smaller wetting angles make the droplet more prone to deformation and elongation. The viscosity ratio also affects the internal vortices of the self-wetting droplet. At higher viscosity ratios, two vortices appear inside the self-rewetting droplet, and the migration speed of the droplet decreases. Besides, the increased viscosity ratio results in greater viscous resistance, which slows down the elongation of the droplet.

\section*{Acknowledgments}

This work was carried out in part using computing resources at the High Performance Computing Platform of Xiangtan University, and it is supported by the National Natural Science Foundation of China (Grant No.  12372287).

%\nolinenumbers
\end{document}